\documentclass[dvipsnames,twocolumn]{aastex631}

\usepackage{placeins}
\usepackage{graphicx}
\usepackage{amsmath}
\usepackage{txfonts}
\usepackage{comment}
\usepackage{color}

\graphicspath{{}}

\newcommand{\mdisk}[0]{\ensuremath{M_{\rm disk}}}
\newcommand{\mgas}[0]{\ensuremath{M_{\rm gas}}}

\newcommand{\rc}[0]{\ensuremath{R_{\rm c}}}

\newcommand{\mstar}[0]{\ensuremath{M_{*}}}
\newcommand{\msun}[0]{\ensuremath{\mathrm{M}_{\odot}}}
\newcommand{\Lsun}[0]{\ensuremath{\mathrm{L}_{\odot}}}
\newcommand{\nhp}[0]{\ensuremath{\mathrm{N}_2\mathrm{H}^+}}

\newcommand{\xco}[0]{\ensuremath{^{13}\mathrm{CO}}}
\newcommand{\cyo}[0]{\ensuremath{\mathrm{C}^{18}\mathrm{O}}}
\newcommand{\zet}[0]{\ensuremath{\zeta_{\rm mid}}}

\newcommand{\abu}[0]{\ensuremath{x_{\rm CO}}}

\begin{document}

\title{ExoALMA XIII. gas masses from \nhp\ and \cyo:\\ a comparison of protoplanetary gas disk mass measurement techniques}

\correspondingauthor{Leon Trapman}
\email{ltrapman@wisc.edu}
\author[0000-0002-8623-9703]{Leon Trapman}
\affiliation{Department of Astronomy, University of Wisconsin-Madison, 
475 N Charter St, Madison, WI 53706, USA}

\author[0000-0003-4663-0318]{Cristiano Longarini}
\affiliation{Institute of Astronomy, University of Cambridge, Madingley Road, Cambridge CB3 0HA, UK}
\affiliation{Dipartimento di Fisica, Università degli Studi di Milano, Via Celoria 16, I-20133 Milano, Italy}

\author[0000-0003-4853-5736]{Giovanni P. Rosotti}
\affiliation{Dipartimento di Fisica, Università degli Studi di Milano, Via Celoria 16, I-20133 Milano, Italy}

\author[0000-0003-2253-2270]{Sean M. Andrews}
\affiliation{Center for Astrophysics | Harvard \& Smithsonian, Cambridge, MA 02138, USA}

\author[0000-0001-7258-770X]{Jaehan Bae}
\affiliation{Department of Astronomy, University of Florida, Gainesville, FL 32611, USA}

\author[0000-0001-6378-7873]{Marcelo Barraza-Alfaro}
\affiliation{Department of Earth, Atmospheric, and Planetary Sciences, Massachusetts Institute of Technology, Cambridge, MA 02139, USA}

\author[0000-0002-7695-7605]{Myriam Benisty}
\affiliation{Univ. Grenoble Alpes, CNRS, IPAG, 38000 Grenoble, France}
\affiliation{Université Côte d'Azur, Observatoire de la Côte d'Azur, CNRS, Laboratoire Lagrange, France}
\affiliation{Max-Planck Institute for Astronomy (MPIA), Königstuhl 17, 69117 Heidelberg, Germany}

\author[0000-0002-2700-9676]{Gianni Cataldi}
\affiliation{National Astronomical Observatory of Japan, 2-21-1 Osawa, Mitaka, Tokyo 181-8588, Japan}

\author[0000-0003-2045-2154]{Pietro Curone}
\affiliation{Dipartimento di Fisica, Università degli Studi di Milano, Via Celoria 16, I-20133 Milano, Italy}
\affiliation{Departamento de Astronomía, Universidad de Chile, Camino El Observatorio 1515, Las Condes, Santiago, Chile}

\author[0000-0002-1483-8811]{Ian Czekala}
\affiliation{School of Physics \& Astronomy, University of St. Andrews, North Haugh, St. Andrews KY16 9SS, UK}
\affiliation{Centre for Exoplanet Science, University of St. Andrews, North Haugh, St. Andrews, KY16 9SS, UK}

\author[0000-0003-4689-2684]{Stefano Facchini}
\affiliation{Dipartimento di Fisica, Università degli Studi di Milano, Via Celoria 16, I-20133 Milano, Italy}

\author[0000-0003-4679-4072]{Daniele Fasano}
\affiliation{Université Côte d'Azur, Observatoire de la Côte d'Azur, CNRS, Laboratoire Lagrange, France}
\affiliation{Univ. Grenoble Alpes, CNRS, IPAG, 38000 Grenoble, France}

\author[0000-0002-9298-3029]{Mario Flock}
\affiliation{Max-Planck Institute for Astronomy (MPIA), Königstuhl 17, 69117 Heidelberg, Germany}

\author[0000-0003-1117-9213]{Misato Fukagawa}
\affiliation{National Astronomical Observatory of Japan, 2-21-1 Osawa, Mitaka, Tokyo 181-8588, Japan}

\author[0000-0002-5503-5476]{Maria Galloway-Sprietsma}
\affiliation{Department of Astronomy, University of Florida, Gainesville, FL 32611, USA}

\author[0000-0002-5910-4598]{Himanshi Garg}
\affiliation{School of Physics and Astronomy, Monash University, VIC 3800, Australia}

\author[0000-0002-8138-0425]{Cassandra Hall}
\affiliation{Department of Physics and Astronomy, The University of Georgia, Athens, GA 30602, USA}
\affiliation{Center for Simulational Physics, The University of Georgia, Athens, GA 30602, USA}
\affiliation{Institute for Artificial Intelligence, The University of Georgia, Athens, GA, 30602, USA}

\author[0000-0001-6947-6072]{Jane Huang}
\affiliation{Department of Astronomy, Columbia University, 538 W. 120th Street, Pupin Hall, New York, NY 10027, USA}

\author[0000-0003-1008-1142]{John D. Ilee}
\affiliation{School of Physics and Astronomy, University of Leeds, Leeds, UK, LS2 9JT}

\author[0000-0001-8446-3026]{Andres F. Izquierdo}
\altaffiliation{NASA Hubble Fellowship Program Sagan Fellow}
\affiliation{Leiden Observatory, Leiden University, P.O. Box 9513, NL-2300 RA Leiden, The Netherlands}
\affiliation{European Southern Observatory, Karl-Schwarzschild-Str. 2, D-85748 Garching bei München, Germany}
\affiliation{Department of Astronomy, University of Florida, Gainesville, FL 32611, USA}

\author[0000-0001-7235-2417]{Kazuhiro Kanagawa}
\affiliation{College of Science, Ibaraki University, 2-1-1 Bunkyo, Mito, Ibaraki 310-8512, Japan}

\author[0000-0002-8896-9435]{Geoffroy Lesur}
\affiliation{Univ. Grenoble Alpes, CNRS, IPAG, 38000 Grenoble, France}

\author[0000-0002-2357-7692]{Giuseppe Lodato}
\affiliation{Dipartimento di Fisica, Università degli Studi di Milano, Via Celoria 16, I-20133 Milano, Italy}

\author[0000-0002-8932-1219]{Ryan A. Loomis}
\affiliation{National Radio Astronomy Observatory, Charlottesville, VA 22903, USA}

\author[0000-0003-4039-8933]{Ryuta Orihara}
\affiliation{College of Science, Ibaraki University, 2-1-1 Bunkyo, Mito, Ibaraki 310-8512, Japan}

\author[0000-0002-4044-8016]{Teresa Paneque-Carreno}
\altaffiliation{51 Pegasi b Fellow}
\affiliation{Department of Astronomy, University of Michigan, 1085 South University Avenue, Ann Arbor, MI 48109, USA}

\author[0000-0001-5907-5179]{Christophe Pinte}
\affiliation{Univ. Grenoble Alpes, CNRS, IPAG, 38000 Grenoble, France}
\affiliation{School of Physics and Astronomy, Monash University, VIC 3800, Australia}

\author[0000-0002-4716-4235]{Daniel Price}
\affiliation{School of Physics and Astronomy, Monash University, VIC 3800, Australia}

\author[0000-0002-0491-143X]{Jochen Stadler}
\affiliation{Université Côte d'Azur, Observatoire de la Côte d'Azur, CNRS, Laboratoire Lagrange, France}
\affiliation{Univ. Grenoble Alpes, CNRS, IPAG, 38000 Grenoble, France}

\author[0000-0003-1534-5186]{Richard Teague}
\affiliation{Department of Earth, Atmospheric, and Planetary Sciences, Massachusetts Institute of Technology, Cambridge, MA 02139, USA}

\author[0000-0002-1284-5831]{Sierk van Terwisga}
\affiliation{Space Research Institute, Austrian Academy of Sciences, Schmiedlstr. 6, A-8042, Graz, Austria}

\author[0000-0003-1859-3070]{Leonardo Testi}
\affiliation{Alma Mater Studiorum Universit\'{a} di Bologna, Dipartimento di Fisica e Astronomia (DIFA), Via Gobetti 93/2, 40129 Bologna, Italy}
\affiliation{INAF-Osservatorio Astrofisico di Arcetri, Largo E. Fermi 5, 50125 Firenze, Italy}

\author[0000-0003-1412-893X]{Hsi-Wei Yen}
\affiliation{Academia Sinica Institute of Astronomy \& Astrophysics, 11F of Astronomy-Mathematics Building, AS/NTU, No.1, Sec. 4, Roosevelt Rd, Taipei 10617, Taiwan}

\author[0000-0002-3468-9577]{Gaylor Wafflard-Fernandez}
\affiliation{Univ. Grenoble Alpes, CNRS, IPAG, 38000 Grenoble, France}

\author[0000-0003-1526-7587]{David J. Wilner}
\affiliation{Center for Astrophysics | Harvard \& Smithsonian, Cambridge, MA 02138, USA}

\author[0000-0002-7501-9801]{Andrew J. Winter}
\affiliation{Université Côte d'Azur, Observatoire de la Côte d'Azur, CNRS, Laboratoire Lagrange, France}
\affiliation{Max-Planck Institute for Astronomy (MPIA), Königstuhl 17, 69117 Heidelberg, Germany}

\author[0000-0002-7212-2416]{Lisa W\"olfer}
\affiliation{Department of Earth, Atmospheric, and Planetary Sciences, Massachusetts Institute of Technology, Cambridge, MA 02139, USA}

\author[0000-00001-8002-8473]{Tomohiro C. Yoshida}
\affiliation{National Astronomical Observatory of Japan, 2-21-1 Osawa, Mitaka, Tokyo 181-8588, Japan}
\affiliation{Department of Astronomical Science, The Graduate University for Advanced Studies, SOKENDAI, 2-21-1 Osawa, Mitaka, Tokyo
45181-8588, Japan}

\author[0000-0001-9319-1296]{Brianna Zawadzki}
\affiliation{Department of Astronomy, Van Vleck Observatory, Wesleyan University, 96 Foss Hill Drive, Middletown, CT 06459, USA}

\author[0000-0002-0661-7517]{Ke Zhang}
\affiliation{Department of Astronomy, University of Wisconsin-Madison, 
475 N Charter St, Madison, WI 53706, USA}

\begin{abstract}
The gas masses of protoplanetary disks are important but elusive quantities. 
In this work we present new ALMA observations of \nhp~(3-2) for 11 exoALMA disks. \nhp\ is a molecule sensitive to CO freeze-out and has recently been shown to significantly improve the accuracy of gas masses estimated from CO line emission. We combine these new observations with archival \nhp\ and CO isotopologue observations to measure gas masses for 19 disks, predominantly from the exoALMA Large program. For 15 of these disks the gas mass has also been measured using gas rotation curves. 
We show that the CO + \nhp\ line emission-based gas masses typically agree with the kinematically measured ones within a factor 3 ($\sim1-2\sigma$). Gas disk masses from CO + \nhp\ are on average a factor $2.3^{+0.7}_{-1.0}\times$ lower than the kinematic disk masses, which could suggest slightly lower N$_2$ abundances and/or  lower midplane ionization rates than typically assumed. Herbig disks are found to have ISM level CO gas abundances based on their CO and \nhp\ fluxes, which sets them apart from T-Tauri disks where abundances are typically $\sim3-30\times$ lower.
The agreement between CO + \nhp-based and kinematically measured gas masses is promising and 
shows that multi-molecule line fluxes are a robust tool to accurately measure disk masses at least for extended disks.

\end{abstract}
%
\section{Introduction}
\label{sec: introduction}

The gas masses of protoplanetary disks are crucial information for planet formation (e.g. \citealt{MorbidelliRaymond2016}). The total mass determines if, and how many, gas giants can be formed in the disk. Furthermore, the gas density plays a large role in the dust physics occurring in the disk, such grain growth and inward drift rates (e.g. \citealt{Drazkowska2023,Birnstiel2024}). The gas surface density also dictates the migration of planets that have already formed (e.g. \citealt{Paardekooper2023}).   

Measuring the gas mass has been plagued with difficulties (see reviews by \citealt{BerginWilliams2018,Miotello2023,Oberg2023}), mainly because H$_2$, the dominant component of the gas, does not significantly emit at the $\sim20-30$ K temperature of the bulk of gas mass in protoplanetary disks (e.g. \citealt{Thi2001,Carmona2011,Pascucci2013,Pinte2018,Law2021bMAPS,Law2022,Paneque-Carreno2023}). 

The gas mass therefore has to be measured indirectly. This is most commonly done from the millimeter (mm) dust continuum emission, where the observed flux is converted into a dust mass by assuming a dust opacity, average disk temperature and that the emission is optically thin (e.g. \citealt{Beckwith1990,Williams2005,ansdell2016,manara2023}). This dust mass is then converted to a gas mass by assuming a gas-to-dust mass ratio, most commonly using the typical interstellar medium (ISM) value of 100. While perhaps the most straightforward way to estimate the disk mass, the dust properties that set the opacity and the gas-to-dust mass ratio are likely different for each disk. Furthermore, there is increasing evidence that some or all of the disk continuum emission is optically thick at wavelengths $\lesssim1.3$mm (e.g. \citealt{Andrews2018b,Tazzari2021a}).

The gas mass can also be inferred from less abundant species in the gas. Among these, hydrogen deuteride (HD) is perhaps the most promising because of its close chemical similarities to H$_2$ (e.g. \citealt{Trapman2017,calahan2021}). \emph{Herschel} detected HD $J=1-0$ ($112 \mu$m) emission towards three disks which allowed for robust gas disk mass measurements (e.g. \citealt{Bergin2013,McClure2016,calahan2021,Schwarz2021,Trapman2022b}), but with the end of the \emph{Herschel} mission we currently lack a far-infrared observatory capable of observing HD in more disks.

The most commonly used gas mass tracer is therefore CO, the second most abundant molecule in the gas, whose bright millimeter lines are detected towards most protoplanetary disks (e.g. \citealt{Dutrey1996,Dent2005,WilliamsBest2014,ansdell2016,ansdell2018,Long2017}). Initial gas mass measurements based on the optically thin isotopologues of CO, \xco\ and \cyo, revealed gas-to-dust mass ratios that were much lower than the ISM value (e.g. \citealt{WilliamsBest2014,miotello2017,Long2017}). Cross comparisons with independent gas mass tracers, such as HD, suggested that these CO-based gas masses were underestimating the true disk mass. 

The likely cause for this underestimation is the omission of one or more processes affecting gaseous CO in disks (e.g. \citealt{Miotello2023}). The original thermochemical models used to derive gas masses included (isotope-selective) photodissociation and freeze-out, the two dominant processes that set the CO abundance structure (e.g. \citealt{Aikawa2002,Visser2009,Miotello2014,Ruaud2019}). Several processes have been suggested to explain the low CO-based gas masses, including the chemical conversion of CO into more complex, less volatile species (e.g., \citealt{Aikawa1997,FuruyaAikawa2014,Yu2016,Yu2017,Bosman2018b,Schwarz2018,Ruaud2022,
Furuya2022}) and locking up CO into larger dust bodies that settle towards the midplane (e.g. \citealt{Bergin2010,Bergin2016,Kama2016,Krijt2018,Krijt2020,
vanClepper2022,Powell2022}).

An alternative approach was suggested by \cite{Anderson2019} and \cite{Trapman2022b}, namely to combine observations of CO and \nhp. In the presence of CO-rich gas the formation of \nhp\ is inhibited and its destruction is increased, making \nhp\ an excellent tracer of CO-poor gas in protoplanetary disks. By simultaneously fitting CO and \nhp, the latter provides an observational estimate of how much CO has been removed from the gas by the processes mentioned previously and therefore how much the CO-based gas mass needs to be corrected to obtain the correct total gas mass.
\cite{Trapman2022b} showed that gas masses measured from the combination of CO isotopologue and \nhp\ lines are consistent with HD-based gas masses. This approach was also recently used by \cite{AGEPRO_V_gas_masses} to measure gas masses for twenty of the disks in the AGE-PRO ALMA Large program \citep{AGEPRO_I_overview}.

The unprecedented spatial and spectral resolution and sensitivity of the Atacama Large Millimeter/submillimeter Array (ALMA) has made it possible to measure protoplanetary disk masses in a new and completely independent way, namely through gas kinematics (for a review \cite{pinte23}). Briefly, while the motions of gas in protoplanetary disks are dominated by the gravitational potential of star modulated by the pressure gradient, the disk self-gravity can leave an observable signature in the gas kinematics from which the disk mass can be measured. In contrast to the methods discussed previously, kinematically measured gas masses trace the total disk mass directly and circumvent the main sources of uncertainty that affect the indirect methods, such as CO chemistry and dust grain properties.

This technique was first applied by \cite{Veronesi2021} for the disk around Elias 2-27 and similar mass measurements have been carried out for the five disks in from the MAPS ALMA Large program (\citealt{Lodato2023,Martire2024}; see also \citealt{Andrews2024,Veronesi2024}).
In an accompanying paper \cite{Longarini2025exoALMA} used the rotation curves derived by \cite{Stadler2025exoALMA} from observations from the exoALMA Large program \citep{Teague2025exoALMA} to kinematically measure disk masses of a subset of the exoALMA sample. They focused their attention on ten disks, for which the geometrical orientation in the sky allows robust constraints on disk masses from self-gravity.

Given the importance of gas disk mass for planet formation, it is crucial that indirect, but more widely applicable, line-flux-based gas masses are compared and benchmarked against kinematically measured gas masses.
In this work we carry out this benchmark and present new \nhp\ $J=3-2$ ALMA Band 7 observations of 11 disks and new \cyo\ $J=2-1$ observations for one disk (RXJ1842.9-3532). We combine these observations with archival ALMA observations of \nhp, \xco\ and \cyo\ to measure gas disk masses for 19 disks, 
which we compare to kinematically based gas masses from the literature \citep{Veronesi2021,Martire2024,Longarini2025exoALMA}.

The structure of this paper is as follows: In Section \ref{sec: observations} we outline our sample and describe the calibration and imaging of the new and archival observations. In Section \ref{sec: methods} we describe how we measure gas mass from these observations. Section \ref{sec: results} analyzes these gas masses and compares them to the kinematically measured gas masses. These results are discussed in Section \ref{sec: discussion} and we present our conclusions in Section \ref{sec: conclusions}.


\begin{table*}[!htbp]
\centering
\caption{\label{tab: exoALMA parameters} Stellar and disk parameters }
\def\arraystretch{1.0}
\begin{tabular*}{1.0\textwidth}{l|cc|ccccc|ccccc|c}
\hline\hline
Name & RA & Dec & dist & SpT & T$_{\rm eff}$ & L$_*$ & M$_*$ & F$_{\rm 1.3mm}$ & M$_{\rm dust}^{\dagger}$ & PA & $i$ & $v_{\rm LSR}$ & Ref.\\
  &  &  & [pc] &  & [K] & [L$_{\odot}$] & [M$_{\odot}$] & [mJy] & [M$_{\oplus}$] & [deg] & [deg] & [km/s] & \\
\hline
DM Tau     & 04:33:48.73 & 18:10:09.97  & 144 & M2    & 3715 & 0.24            & 0.45 & 104.7 & 79.3  & 336 & 39 & 6.0  & 1,2 \\
AA Tau     & 04:34:55.42 & 24:28:53.03  & 135 & K7    & 4350 & 1.10            & 0.79 & 67.0  & 27.8  & 273 & 59 & 6.4  & 6 \\ 
LkCa 15    & 04:39:17.79 & 22:21:03.39  & 157 & K5    & 4365 & 1.00            & 1.14 & 151.0 & 85.9  & 62  & 50 & 6.2  & 5 \\ 
HD 34282   & 05:16:00.48 & -09:48:35.39 & 309 & A0    & 9520 & 10.8            & 1.61 & 99.0  & 109.4 & 117 & 58 & -2.4 & 11,13 \\ 
MWC 758    & 05:30:27.53 & 25:19:57.08  & 156 & A7    & 7850 & 14.0            & 1.40 & 70.7  & 19.1  & 240 & 19 & 6.0  & 13,14 \\ 
CQ Tau     & 05:35:58.47 & 24:44:54.09  & 149 & F2    & 6890 & 10.0            & 1.4 & 144.0 & 44.4  & 235 & 52 & 6.1   & 12,13 \\ 
PDS 66     & 13:22:07.54 & -69:38:12.22 & 98  & K1V   & 5035 & 1.20            & 1.28 & 148.0 & 30.4  & 189 & 32 & 4.0  & 7,8 \\ 
HD135344B  & 15:15:48.45 & -37:09:16.02 & 135 & F5    & 6440 & 6.70            & 1.61 & 117.0 & 28.1  & 243 & 16 & 7.0  & 11 \\ 
HD143006   & 15:58:36.91 & -22:57:15.22 & 167 & G7    & 5620 & 3.80            & 1.56 & 59.0  & 24.6  & 169 & 17 & 7.8  & 9,10 \\ 
RXJ1615.3-3255  & 16:15:20.23 & -32:55:05.10 & 156 & K5    & 4350 & 0.60            & 1.14 & 480.0 & 312.3 & 325 & 47 & 8.0  & 1 \\ 
V4046 Sgr  & 18:14:10.48 & -32:47:34.52 & 72  & K5/K7 & 4370 & 0.5/0.3$^{*}$   & 1.73 & 327.0 & 47.6  & 256 & 34 & 3.0  & 3 \\ 
RXJ1842.9-3532  & 18:42:57.98 & -35:32:42.83 & 151 & K2    & 4780 & 0.80            & 1.07 & 63.2  & 36.6  & 206 & 39 & 6.0  & 4 \\ 
RXJ1852.3-3700  & 18:52:17.30 & -37:00:11.95 & 147 & K2    & 4780 & 0.60            & 1.03 & 58.2  & 33.2  & 147 & 33 & 5.6  & 4 \\ 
GM Aur     & 04:55:10.98 & 21:59:54.00  & 159 & K6    & 4350 & 1.20            & 1.10 & 162.0 & 88.3  & 57  & 53 & 5.6  & 15,16 \\ 
AS 209     & 16:49:15.30 & -14:22:08.64 & 121 & K5    & 4266 & 1.41            & 1.20 & 288.0 & 86.5  & 86  & 35 & 4.6  & 17,18 \\ 
IM Lup     & 15:56:09.21 & -37:56:06.13 & 158 & K5    & 4266 & 2.57            & 1.10 & 253.0 & 108.3 & 144 & 48 & 4.5  & 19,20 \\ 
MWC 480    & 04:58:46.30 & 29:50:37.00  & 162 & A5    & 8250 & 21.9            & 2.10 & 253.0 & 61.9  & 148 & 37 & 5.1  & 21 \\ 
HD 163296  & 17:56:21.30 & -21:57:22.00 & 101 & A1    & 9355 & 17.0            & 2.00 & 715.0 & 73.0  & 133 & 47 & 4.8  & 11 \\ 
Elias 2-27 & 16:26:45.0 & -24:23:7.8    & 116 & M0    & 3850 & 0.92            & 0.49 & 330.0 & 103.9 & 119 & 56 & 2.0  & 18 \\ 
\hline\hline
\end{tabular*}
\begin{minipage}{0.93\textwidth}
\vspace{0.1cm}
{\footnotesize{\textbf{Notes:}$^*$V4046 Sgr is a compact binary system. The table lists the spectral type
 and stellar luminosity for both its stars. $^{\dagger}$ Dust masses are computed from the 1.3 millimeter continuum flux using 
 M$_{\rm dust}$ = F$_{\rm 1.3mm}d^2/\kappa_{\nu}B_{\nu}$(T) assuming $\kappa_{\nu} = 2.3\times(\nu/{\rm 230~GHz})~{\rm cm}^2~{\rm g}^{-1}$ and T = 20 K (cf. \citealt{Beckwith1990,manara2023}). \textbf{References:} $\mstar, i$ and PA from \cite{Izquierdo_exoALMA}, R.A., Dec and dist from Table 1 in \cite{Teague2025exoALMA}. 
 1: \citealt{Herbig1977},
 2: \citealt{Andrews2011},
 3: \citealt{Bouvier1999},
 4: \citealt{Andrews2018b},
 5: \citealt{Fairlamb2015},
 6: \citealt{Guzman-Diaz2021},
 7: \citealt{Acke2005},
 8: \citealt{Herbig1960},
 9: \citealt{Mamajek2002},
 10: \citealt{Ribas2023},
 11: \citealt{LuhmanMamajek2012},
 12: \citealt{Barenfeld2016},
 13: \citealt{Quast2000},
 14: \citealt{Manara2014},
 15: \citealt{Espaillat2010},
 16: \citealt{Macias2018},
 17: \citealt{HerbigBell1988},
 18: \citealt{Andrews2009},
 19: \citealt{alcala2017},
 20: \citealt{Alcala2019},
 21: \citealt{Montesinos2009} 
}}
\end{minipage}
\end{table*}

\section{Observations \& Sample}
\label{sec: observations}

\subsection{Sample}
\label{sample}

The sample discussed in this work differs somewhat from the exoALMA disk sample \citep{Teague2025exoALMA}. Of the 15 exoALMA disks, all but two, SY Cha and RXJ1604-3010 A, have observations of \nhp~(3-2) and \xco\ and \cyo~(2-1) or (3-2) available in the ALMA archive or presented in this work. As we will discuss in Section \ref{sec: methods} these lines are needed to measure the gas mass, so we limit our analysis to these 13 disks. For nine of the 13 disks \citet{Longarini2025exoALMA} were able to measure kinematic gas masses from the rotation curves presented in \citet{Stadler2025exoALMA}. The remaining four disks were either too face-on or their emission was too asymmetric to obtain reliable kinematic gas masses (see \citealt{Longarini2025exoALMA} for details). While this prevents us from performing a gas mass comparison with kinematics, we do include these four disks in our other analyses.

We supplement our 13 exoALMA disks with six disks where kinematically measured gas masses are also available \citep{Veronesi2021,Martire2024} to maximize the number of disks with such measurements in our sample. These are the five disks from the MAPS large program \citep{Oberg2021MAPS} and Elias 2-27 (e.g. \citealt{Perez2016,Paneque-Carreno2021}). These six disks also have the necessary line fluxes available to measure their gas masses. Our full sample thus consists of 19 disks, 15 of which have kinematically measured gas masses. Stellar and disk parameters of the sample are summarized in Table \ref{tab: exoALMA parameters}.

\subsection{New observations}
\label{sec: new observations}

We present new ALMA Band 7 observations of \nhp\ $J=3-2$ for AA Tau, CQ Tau, MWC 758, RXJ1615.3-3255, RXJ1852.3-3700, RXJ1842.9-3532, PDS 66, HD 34282, HD 143006, HD 135344B, and Elias 2-27 (IDs 2022.1.00485.S, 2023.1.00334.S; PI: L. Trapman). We also present ALMA Band 6 observations of \cyo~(2-1) and C$^{17}$O 2-1 for RXJ1842.9-3532 (ID 2023.1.00334.S). 

The ALMA Band 6 and 7 observations were carried out October 2023-May 2024 for program 2023.1.00334.S, with baseline lengths between 15 and 1200 m and October 2022-January 2023 for program 2022.1.00485.S, with baseline lengths between 15 and 740 m. 

The Band 7 correlator was set up to cover \nhp~(3-2) at 279.517 GHz, DCO$^{+}$ 4-3 at 288.149 GHz, DCN 4-3 at 289.65 GHz, C$^{34}$S 6-5 at 189.215 GHz and H$_2$CO 4$_{0,4}$ - 3$_{0,3}$ at 290.623 GHz. All lines are covered at a spectral resolution of 122 kHz ($\sim 120 $ m/s) with the exception of H$_2$CO, which was covered at a coarser spectral resolution of 977 kHz ($1.2$ km/s). In addition to the lines the correlator setup also includes two 1.875 GHz wide continuum spectral window centered at 278.008 GHz and 290.870 GHz, respectively. Note that the higher frequency continuum spectral window is the same spectral window that covers the H$_2$CO line, hence why it has a coarser spectral resolution.

The Band 6 correlator was set up to cover the \cyo~(2-1) line at 219.565 GHz, C$^{17}$O 2-1 at 224.719 GHz, SO$_2$ 4$_{2,2}$ - 3$_{1,3}$ at 235.156 GHz, HCC$^{13}$CN 26-25 at 235.514 GHz, HC$^{13}$CCN 26-25 at 235.537 GHz, and HCCCN 26-25 at 236.517 GHz. All lines are covered at a spectral resolution of 122 kHz ($\sim150$ m/s) with the exception of \cyo~(2-1), which is covered at a higher spectral resolution of 61 kHz (83.3 m/s). In addition to the lines the correlator setup also includes a 1.875 GHz wide continuum spectral window centered at 237.063 GHz.

The on-source times for Band 7 and Band 6 were 30-50 min and 40 min, respectively. 

\begin{figure*}[ht]
    \centering
    \includegraphics[width=\textwidth]{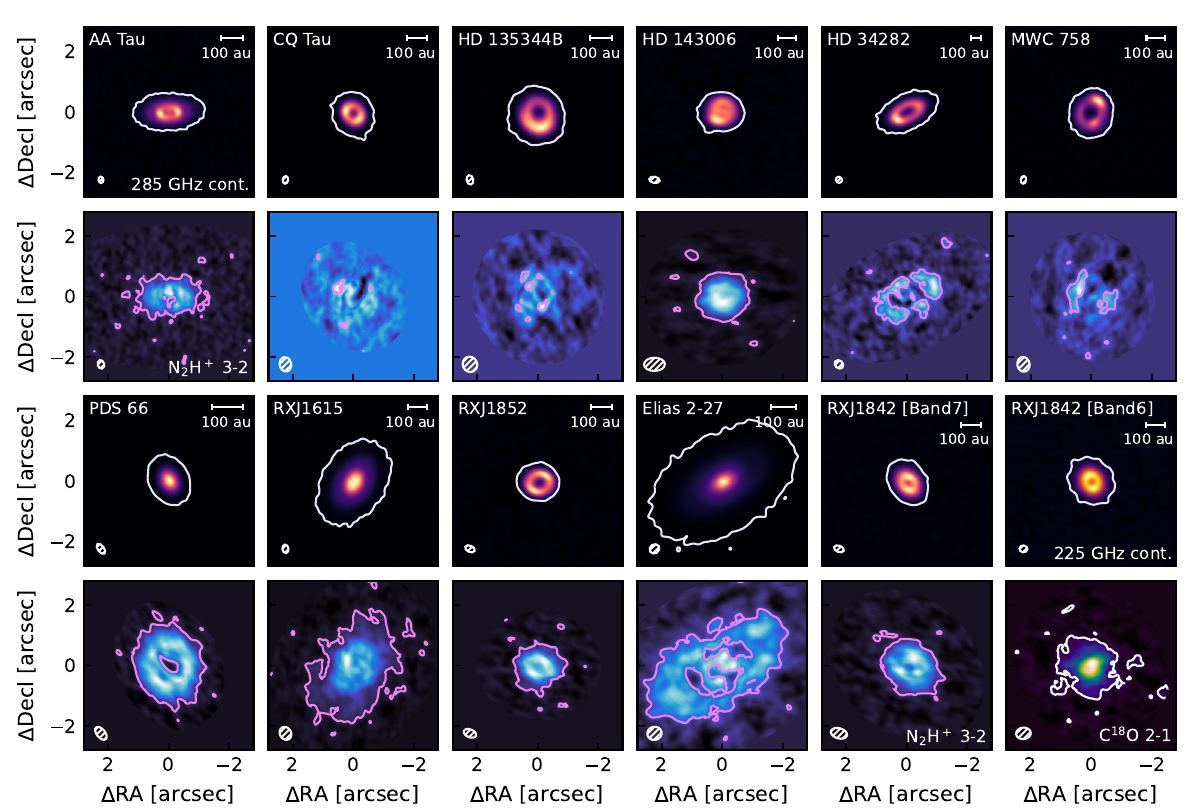}
    \caption{\label{fig: B7 obs gallery} Gallery of the new ALMA Band 6 and 7 observations presented this work. Colors are peak-normalized for each individual panel. For each source the 285 GHz continuum is shown first, with its \nhp~(3-2) integrated intensity map shown directly below that. Note that the last two panels show the Band 6 225 GHz and \cyo~(2-1) observations of RXJ1842.9-3532. White (except for \cyo) and pink contours denote $5\sigma$ and $3\sigma$, respectively. The white contours for \cyo\ also shows $3\sigma$. }
\end{figure*}

\subsubsection{Data reduction}
\label{sec: data reduction}

All observations were initially calibrated using the Common Astronomy Software Application (CASA; \citealt{CASAteam2022}) pipeline version 6.4.1.12 for program 2022.1.00485.S and 6.5.4.9 for program 2023.1.00334.S using the \texttt{scriptforPI.py} as provided with the data.

The signal-to-noise (SNR) of our observations was high enough to conduct self-calibration after the initial pipeline calibration. Our approach was based on the self-cal procedures of the DSHARP, MAPS, and AGE-PRO large programs (e.g. \citealt{Andrews2018,Czekala2021MAPS,AGEPRO_I_overview}). It is near-identical to the exoALMA self-calibration procedure, with the exception of the spatial alignment (see \citealt{Loomis_exoALMA} for details). 

First, continuum-only visibility data sets were created by flagging all channels with $\pm20$ km/s of one of the lines in our spectral setup, irrespective of whether this line was detected or not. The resulting line-free continuum channels were averaged together to channels with a maximum width of 125 MHz. Next, individual execution blocks were astrometrically aligned
by fitting 2D Gaussians to their continuum emission in the image plane and aligning their phase centers to the coordinates listed in Table \ref{tab: exoALMA parameters}. For HD135344B, HD143006, HD34282, and MWC 758 the continuum emission was found to be too asymmetric to be reliably fitted with a Gaussian. Here we instead fitted the data with a Gaussian ring, using its center to align the measurement sets.

After alignment we checked relative flux scaling between execution blocks by deprojecting the continuum visibilities using disk geometry from the literature (see Table \ref{tab: exoALMA parameters}) and comparing their amplitude profile as a function of baseline length. In all cases we found the fluxes of individual execution blocks for a single source to differ by less than 10\%. As this is within the expected absolute flux uncertainty of ALMA we did not rescale the fluxes.

Self-calibration was then carried out first on just the short baseline observations, followed by a second self-calibration using the combined self-calibrated short baseline and non-self-calibrated long baseline observations. For both self-calibration sets we performed 1-3 rounds of phase-only self-calibration followed by a single round of amplitude+phase self-calibration. For the phase-only rounds the initial solution interval was chosen to be the full on-source time. Each consecutive round reduced that interval by a factor two, ending when the current phase-only solution improved the peak SNR by less than 10\% compared to the previous solution.

After self-calibration the obtained gaintables were applied to the full visibility dataset, followed by a subtraction of the continuum using the task \texttt{uvcontsub}.
For each spectral window we flag all channels within $\pm 20$ km/s of each line in the spectral setup before fitting the remaining channels with a first-order polynomial.

\subsubsection{Imaging}
\label{sec: imaging}

The lines and continuum were imaged with the \texttt{tclean} CASA task
using the `multiscale' deconvolver with scales of [0, 5, 15] $0\farcs02$ pixels, a `Briggs' weighting with a robust parameter of 0.5, a gain of 0.08 and a small scale bias of 0.45. The images were cleaned down to a threshold of $1\sigma$, measured from an emission-free part of the cube.
For imaging the lines we also include a Keplerian mask created using the routine from \cite{teague2020} with the parameters listed in Table \ref{tab: exoALMA parameters}. This results in a typical beam of $\sim0\farcs25-0\farcs35$ and sensitivities of $\sim3-5$ mJy/beam per 0.25 km/s channel.

\subsection{Archival observations}
\label{sec: archival observation}

In addition to the new \nhp~(3-2) observations presented in this work we also make use of a large set of archival observations, primarily ones that provide \xco\ and \cyo\ observations for our sources. In most cases these are of the $J=2-1$ transition, but for RXJ1615.3-3255, AA Tau, HD 143006, and HD 135344B, observations of this transition were not available in the ALMA archive so the $J=3-2$ transition is used instead. For Elias 2-27 observations of $J=2-1$ are available but show significantly more cloud contamination than the $J=3-2$ observations. We therefore opt to use the latter transition here.
Table \ref{tab: fluxes and programs} summarized the program IDs of the archival data sets and for further details on the observations and data reduction of each we refer the reader to the works listed in the table. 

For two sources, RXJ1615.3-3255 and HD 143006, one or both of the CO isotopologue archival observations have not previously appeared in the literature. For these sources we calibrated, self-calibrated and imaged the archival data following the same procedure as was used for the new observations. A more in depth analysis of the CO isotopologue observations of HD 143006 will be presented in \emph{van Terwisga et al., in prep.} 

\begin{figure}[!htbp]
    \centering
    \begin{minipage}{0.96\columnwidth}
    \includegraphics[width=\textwidth]{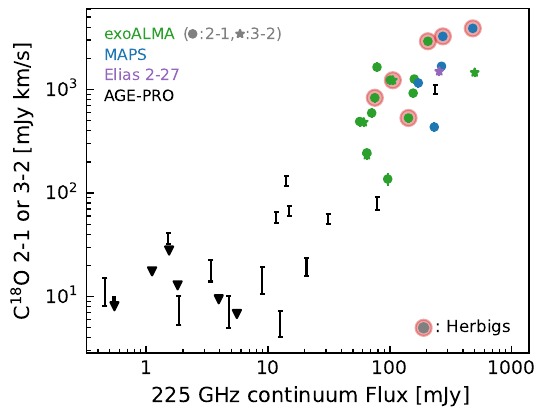}    
    \end{minipage}
    \begin{minipage}{0.96\columnwidth}
    \includegraphics[width=\textwidth]{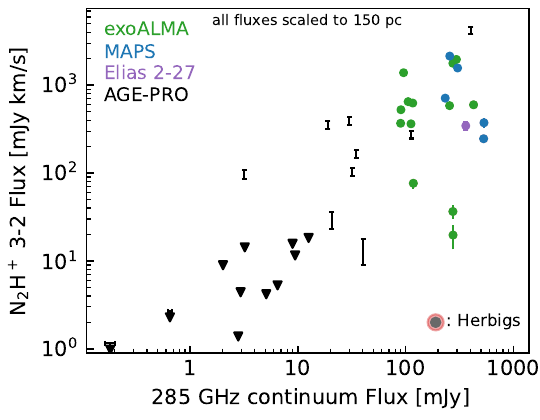}    
    \end{minipage}
    \begin{minipage}{0.96\columnwidth}
    \includegraphics[width=\textwidth]{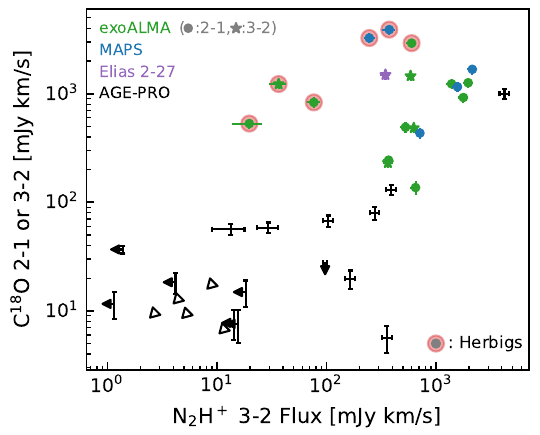}    
    \end{minipage}
    \caption{\label{fig: integrated line fluxes} Integrated line and continuum fluxes, scaled to a distance of 150 pc, of the disk sample examined in this work (in green, blue, and purple) combined with twenty class II disks from the AGE-PRO ALMA large program, shown in black \citep{AGEPRO_III_Lupus,AGEPRO_IV_UpperSco}. \cyo~(2-1) and (3-2) observations are denoted as circles and stars, respectively. The \cyo~(3-2) fluxes are reduced by a factor 2.3, the median \cyo~(3-2)/\cyo~(2-1) line ratio of models with $\mdisk\geq10^{-4}~\msun$ from \cite{AGEPRO_V_gas_masses}. Herbig disks are highlighted with a pink circle. Triangles show upper limits, with open triangles representing upper limits in both \nhp\ and \cyo.  }
\end{figure}

\subsection{A first look at the observations}
\label{sec: first look at the observations}

Integrated intensity maps of the 285 GHz continuum and \nhp\ of each source in our sample are presented in Figure \ref{fig: B7 obs gallery}. While some of the disks exhibit substructures, this work will focus on the gas mass derived from integrated fluxes and the analysis of the resolved emission will be presented in a future work. 

To measure integrated fluxes, the data cubes were first corrected for the non-Gaussian beam resulting from multi-configuration ALMA observations following \citealt{Czekala2021MAPS} (so-called JvM correction, see also \citealt{JorsatervanMoorsel1995}). In the case of line emission, integrated intensity maps were created using the same Keplerian mask used for imaging. From the resulting 2D images the integrated flux was measured using a curve-of-growth method\footnote{as implemented in \url{https://zenodo.org/records/14973489}}. Here the flux is measured using increasingly larger elliptical apertures with the same aspect ratio and orientation as the disk until integrated flux plateaus. 
We note that this method was only used to measure fluxes from the new observations and the archival observations where integrated fluxes had not been previously published. For sources where CO isotopologue or \nhp\ integrated line fluxes were already available we adopt the published values to maintain consistency with existing literature.
The integrated line and continuum fluxes are summarized in Table \ref{tab: fluxes and programs}. 

Figure \ref{fig: integrated line fluxes} shows the integrated line and continuum fluxes from this work combined with the 20 class II disks from the AGE-PRO ALMA large program first presented in \cite{AGEPRO_III_Lupus,AGEPRO_IV_UpperSco}. 
Note that for \cyo\ we have scaled down the \cyo~(3-2) line fluxes by a factor 2.3, which is the median \cyo~(3-2)/\cyo~(2-1) flux line ratio for thermochemical models with $\mdisk \geq 10^{-4}~\msun$ from \cite{AGEPRO_V_gas_masses}.
Starting from the top panel of Figure \ref{fig: integrated line fluxes}, we see a correlation between \cyo\ versus 225 GHz continuum fluxes, which is in line with earlier works (e.g., \citealt{Bergner2019,Anderson2024,AGEPRO_III_Lupus,AGEPRO_IV_UpperSco}). The combined sample of the three ALMA large programs, MAPS, AGE-PRO, and exoALMA, shows that this correlation holds over over three orders of magnitude in both continuum and \cyo\ flux.

If the Herbigs disks are excluded, a similar positive correlation can be seen between \nhp~(3-2) and the 285 GHz continuum fluxes for the T-Tauri disks, also in line with previous findings (e.g., \citealt{Anderson2022,Anderson2024,AGEPRO_III_Lupus,AGEPRO_IV_UpperSco}). It is unclear if the same correlation is also present for faint ($\lesssim 10$ mJy) continuum sources due to the low detection rate of \nhp~(3-2) below that point. It should also be noted that, while suggestive, the trend of \nhp\ upper limits with continuum flux is an artifact of the correlation between continuum flux and continuum disk size (e.g. \citealt{Tripathi2017,Hendler2020}). For the \nhp\ non-detections the continuum dust disk size was used as the aperture size for many sources, which when combined with the continuum flux-size relation leads to a positive correlation between the \nhp\ upper limit threshold and the continuum flux (see \citealt{AGEPRO_III_Lupus,AGEPRO_IV_UpperSco} for details).  

The middle panel of Figure \ref{fig: integrated line fluxes} also shows that the Herbig disks in the sample do not follow the same \nhp-continuum flux correlation as is seen for the T-Tauri disks. For Herbig disks of a given continuum flux the \nhp~(3-2) integrated flux is approximately an order of magnitude fainter. A likely cause for this are the higher stellar luminosities of these sources warming up the disk. This results in less CO freeze-out and therefore a higher CO abundance that inhibits the formation of \nhp\ (e.g. \citealt{Qi2015}). 
Curiously, the \nhp\ flux of Elias 2-27, itself a M0 T-Tauri star, lies closer to the Herbig disks than to the T-Tauri disks. 
A possible explanation could be the young age of the system, with active accretion through the disk and backwarming by an envelope increasing the disk temperature to similar levels as in the Herbig disks (e.g. \citealt{DAlessio1998,Whitney2013,vtHoff2018,vtHoff2020,Kuznetsova2022}).
An alternate explanation could be a low midplane ionization rate in this source, which we will discuss further in Section \ref{sec: constraining midplane ionization rates}.

The bottom panel of Figure \ref{fig: integrated line fluxes} shows \cyo\ set against \nhp. Again excluding the Herbig disks, we see that above a \cyo\ flux of $\sim100$ mJy km/s the two lines are correlated, similar to what was recently found by \cite{Anderson2024}. As discussed by these authors this correlation is unexpected given the anti-correlation between CO and \nhp, but it could suggest that other factors, such as disk mass and size dominate over chemical effects. Alternatively, a correlation would also be expected if CO abundances are similar between disks (e.g. \citealt{AGEPRO_III_Lupus,AGEPRO_IV_UpperSco}).
It is unclear if the correlation extends below this threshold. \nhp\ is only detected for six of the eighteen sources with a \cyo\ flux $\leq100$ mJy km/s, but one of those detections is for the source with the faintest \cyo\ detection in the figure. As these are all AGE-PRO sources we refer the reader to \cite{AGEPRO_III_Lupus,AGEPRO_IV_UpperSco} for a more detailed analysis.

As with the middle panel, the Herbig disks and Elias 2-27 occupy a different part of the figure. These sources have \nhp\ fluxes that are $\sim1-2$ orders of magnitude lower compared to T-Tauri disks with similar \cyo\ fluxes. Located in between the Herbigs and the T-Tauris is the disk around HD 143006, a G7 spectral type star whose stellar luminosity lies in between the T-Tauri and Herbig stars that make up our sample. Its location in the bottom panel of Figure \ref{fig: integrated line fluxes} is consistent with the idea that a higher overall disk temperature, driven by a higher stellar luminosity, is what sets apart Herbig and T-Tauri disks. 

\begin{figure*}[!ht]
    \centering
    \includegraphics[width=\textwidth]{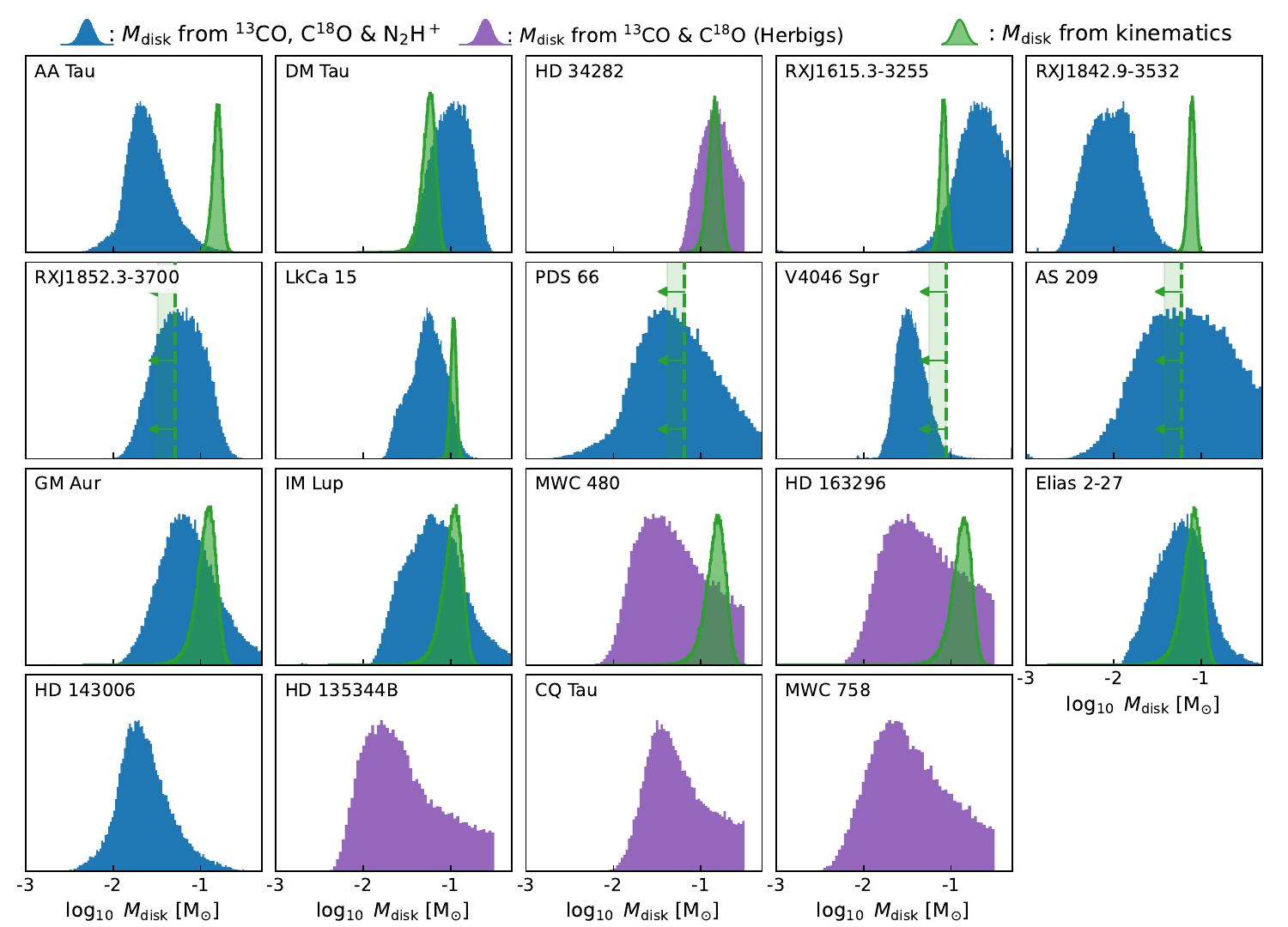}
    \caption{\label{fig: disk mass comparisons} Gas disk mass posteriors obtained from either fitting \xco, \cyo, \nhp\ and the 225 GHz continuum for the T-Tauri disks (shown in blue) or \xco, \cyo\ and the 225 GHz continuum for the Herbig disks (shown in purple). 
    In green we show the posteriors of gas disk masses derived from the disk kinematics by \cite{Veronesi2021,Martire2024,Longarini2025exoALMA}. Gas mass upper limits from kinematics, taken to be 5\% of the stellar mass (e.g. \citealt{Andrews2024,Veronesi2024}), are shown as a green vertical dashed line.}
\end{figure*}

\section{Measuring gas disk masses from line emission}
\label{sec: methods}

For measuring the gas disk masses (\mgas) in our sample we follow the approach from \cite{AGEPRO_V_gas_masses} using the grid of thermochemical models
presented in the same work. To briefly summarize this approach: 
\cite{AGEPRO_V_gas_masses} used the thermochemical code \texttt{DALI} \citep{Bruderer2012,Bruderer2013} to run a large grid of disk models where they varied the disk mass, characteristic size, stellar luminosity, vertical structure, and dust properties. These models include isotope-selective photo-dissociation and chemistry for CO following \citealt{Miotello2014}, but they do not include the processes thought to be responsible for removing CO from the gas in the CO emitting layer. Instead, in their models the peak CO abundance in the CO emitting layer is included as a free parameter ($3\times10^{-7}~\leq x_{\rm CO}\leq 10^{-4}$), which serves as a proxy for processes beyond photodissocation and freeze-out that affect gaseous CO in the disk. Constraining this parameter is the main reason for including \nhp\ in addition to the CO lines, as it breaks the degeneracy between disk mass and $x_{\rm CO}$ encountered when only using CO lines (e.g. \citealt{Anderson2019,Anderson2022,Trapman2022b,Oberg2023}).

Using synthetic observations from this model grid they used a MCMC to carry out a $\chi^2$ fit of the observed 225 GHz continuum, \xco~(2-1), \cyo~(2-1), and \nhp~(3-2) integrated line fluxes and derive a posterior distribution for the disks in AGE-PRO (see \citealt{AGEPRO_V_gas_masses} for further details).

In this work we make a few small modifications to this approach. First, we use $ 10^{-4}\ \msun \leq \mgas \leq 0.5\ \msun$ as a prior for the disk mass, which is in line with the fact that the sources in this work are much brighter, and likely more massive than the AGE-PRO disks. This prior is assumed to be uniform in logspace as the gas mass is also fit in logspace. Second, six T-Tauri sources in our sample have stellar luminosities $L_* \geq 1\ \mathrm{L}_{\odot}$, which is the maximum stellar luminosity in the \citet{AGEPRO_V_gas_masses} model grid. As fitting the observations with a low stellar luminosity could lead to an overestimation of the amount of CO freeze-out and thus an overestimation of the gas disk mass, we ran an extra series of models\footnote{The fluxes for these models and the ones presented in Section \ref{sec: effect inner cavity} and \ref{sec: n2hp for herbigs} are available at Zenodo DOI: \url{https://zenodo.org/records/15148628}} with $L_* = 3.0$, varying all other parameters as in \citet{AGEPRO_V_gas_masses}, except that we skip $R_{\rm c} =1~\mathrm{au}$ and $\mgas < 10^{-4}~\msun$ as these models would be too compact and/or low-mass to represent the disks in our sample. The fluxes for these models and the ones that will be discussed in Sections \ref{sec: effect inner cavity} and \ref{sec: n2hp for herbigs} will be made available as part of this manuscript.

It is also worth pointing out that many of the sources in the sample have large dust cavities and/or other dust substructures (e.g. Figure \ref{fig: B7 obs gallery}; \citealt{Perez2015,Paneque-Carreno2021,Sierra2021,Curone2025exoALMA}), which are not present in thermochemical models. We will revisit their impact on our results in Section \ref{sec: effect inner cavity}.

The Herbig disks in our sample, i.e., HD 135344B, HD 34282, CQ Tau, MWC 758, HD 163296, and MWC 480, warrant a slightly different approach. The models in \cite{AGEPRO_V_gas_masses} were run using stellar properties typical for T-Tauri stars (T$_{\rm eff} = 3500-4000$ K; $L_* = 0.1 - 1.0$ L$_{\odot}$), making them not representative for these sources. Recently, \cite{Stapper2024} presented a large grid of thermochemical models for Herbig disks, which they used to measure gas masses for a large sample of Herbig disks. In order to have a gas mass measurement method as consistent as possible for both T-Tauri and Herbig disks we use the model grid from \cite{Stapper2024} and fit the observations, in this case the \xco, and \cyo, using a similar MCMC as was used for the T-Tauri disks. 
For consistency we also compare our derived Herbig disk gas masses with the values obtained by \cite{Stapper2024}, finding excellent agreement between the two gas mass estimates, with gas masses being within a factor $\sim3$ ($1\sigma$) from each other. 

It should be noted here that the models from \cite{Stapper2024} do not include the peak CO abundances as a free parameter. Several studies find that CO in Herbig disks is underabundant by an order of magnitude (e.g. \citealt{PanequeCarreno2024,Rosotti_exoALMA}). However, as we will show in Section \ref{sec: n2hp for herbigs}, including the CO abundance as a free parameter and fitting \nhp\ in addition to the CO lines does not significantly change the derived gas mass.


\section{Results}
\label{sec: results}

\subsection{Line based gas disk masses}
\label{sec: line based gas disk masses}

\begin{table}[htb]
\centering
\caption{\label{tab: gas masses sample} Emission line based gas masses }
\def\arraystretch{1.05}
\begin{tabular*}{0.9\columnwidth}{l|ccc}
\hline\hline
name  & M$_{\rm gas}$ & $\log_{10}~\mathrm{x}_{\rm CO}$ & $\log_{10}~\zeta_{\rm mid}$ \\
 & [M$_{\odot}$] & [-] & [s$^{-1}$] \\
\hline
 DM Tau &$0.102^{+0.064}_{-0.042}$ &$-4.24^{+0.15}_{-0.21}$ &$-17.75^{+0.68}_{-0.82}$\\
 V4046 Sgr &$0.035^{+0.018}_{-0.010}$ &$-4.60^{+0.20}_{-0.21}$ &$-17.62^{+0.91}_{-0.97}$\\
 RXJ1615.3-3255 &$0.207^{+0.148}_{-0.090}$ &$-4.12^{+0.08}_{-0.12}$ &- \\
 RXJ1852.3-3700 &$0.057^{+0.052}_{-0.027}$ &$-4.54^{+0.22}_{-0.24}$ &$-16.94^{+0.64}_{-0.98}$\\
 RXJ1842.9-3532 &$0.009^{+0.008}_{-0.004}$ &$-4.60^{+0.21}_{-0.22}$ &- \\
 LkCa 15 &$0.055^{+0.031}_{-0.022}$ &$-4.46^{+0.20}_{-0.21}$ &$-18.66^{+1.03}_{-0.84}$\\
 AA Tau &$0.024^{+0.020}_{-0.009}$ &$-4.60^{+0.24}_{-0.22}$ &$-17.71^{+0.98}_{-1.21}$\\
 PDS 66 &$0.050^{+0.102}_{-0.031}$ &$-5.53^{+0.28}_{-0.29}$ &$-17.55^{+0.96}_{-0.94}$ \\
 HD 143006 &$0.021^{+0.024}_{-0.009}$ &$-5.11^{+0.25}_{-0.23}$ &- \\
 HD 135344B &$0.024^{+0.071}_{-0.014}$ &$-4.13^{+0.10}_{-0.17}$ &- \\
 CQ Tau &$0.049^{+0.092}_{-0.024}$ &$\mathit{-4.50^{+0.30}_{-0.08}}$$^{(\dagger)}$ &- \\
 HD 34282 &$0.148^{+0.082}_{-0.052}$ &$-4.11^{+0.07}_{-0.12}$ &- \\
 MWC 758 &$0.031^{+0.076}_{-0.018}$ &$-4.25^{+0.15}_{-0.21}$ &- \\
 GM Aur &$0.069^{+0.072}_{-0.029}$ &$-4.38^{+0.21}_{-0.23}$ &$-18.66^{+0.81}_{-0.77}$\\
 AS 209 &$0.087^{+0.132}_{-0.055}$ &$-4.68^{+0.25}_{-0.26}$ &$-16.80^{+0.54}_{-0.85}$\\
 IM Lup &$0.041^{+0.037}_{-0.018}$ &$-4.45^{+0.21}_{-0.20}$ &$-18.52^{+0.70}_{-0.79}$\\
 MWC 480 &$0.066^{+0.057}_{-0.020}$ &$-4.22^{+0.15}_{-0.21}$ &- \\
 HD 163296 &$0.074^{+0.078}_{-0.030}$ &$-4.28^{+0.17}_{-0.21}$ &- \\
 Elias 2-27 &$0.060^{+0.053}_{-0.030}$ &$-4.20^{+0.14}_{-0.19}$ &$-19.48^{+0.41}_{-0.33}$\\
\hline\hline
\end{tabular*}
\begin{minipage}{0.93\columnwidth}
\vspace{0.1cm}
{\footnotesize{\textbf{Notes:}
Best fit gas disk masses are the median of \mgas-posteriors presented in Figures \ref{fig: disk mass comparisons}
, with the uncertainties being the 16$^{\rm th}$ and 84$^{\rm th}$ quantile of the distribution. Herbig disk CO abundances are from
 the models discussed in Section \ref{sec: n2hp for herbigs}. Midplane ionization rates are from the fits discussed
 in Section \ref{sec: constraining midplane ionization rates}.$^{(\dagger)}$: The $\mathrm{x}_{\rm CO}$ fit for CQ Tau is bimodal and this value should be viewed with caution (see Section \ref{sec: n2hp for herbigs} for details).
}}
\end{minipage}
\end{table}

 Figure \ref{fig: disk mass comparisons} shows the gas masses derived from fitting the 225 GHz dust continuum and \xco, \cyo, and \nhp\ integrated line fluxes. The gas masses are also summarized in Table \ref{tab: gas masses sample}. We find overall high gas masses in the range of $\sim0.01 - 0.1\ \msun$, which is unsurprising as the disks in the sample are among the brightest, and therefore most massive and/or largest, protoplanetary disks in the Solar neighborhood (e.g. \citealt{Oberg2021MAPS,Teague2025exoALMA}).

\begin{figure}[htb]
    \centering
    \includegraphics[width=0.98\columnwidth]{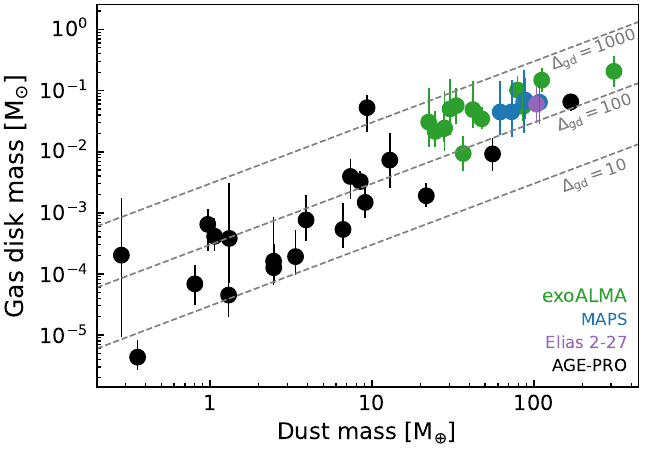}
    \caption{\label{fig: mgas versus mdust} Emission line based gas disk masses versus dust masses for the 19 sources examined in this work plus the 20 class II disks from the AGE-PRO ALMA large program \citep{AGEPRO_I_overview,AGEPRO_III_Lupus,AGEPRO_IV_UpperSco,AGEPRO_V_gas_masses}. 
    Note that the 1.3 mm flux used to derive the dust mass is very likely at least somewhat optically thick for the brightest sources, so dust masses $\gtrsim20~\mathrm{M}_{\oplus}$ should be considered lower limits of the true dust mass. Gray dashed lines show gas-to-dust mass ratios of 10, 100, and 1000.  }
\end{figure}

In Figure \ref{fig: mgas versus mdust} we plot \mgas\ against the dust mass, computed from the continuum flux assuming a disk averaged temperature of 20 K and optically thin emission (see also \citealt{Hildebrand1983,ansdell2016,manara2023}). We find gas-to-dust mass ratios between $\sim80$ and $\sim530$ for the 19 sources examined here, similar to what is found by \citet{Longarini2025exoALMA} using the kinematically measured gas masses and dust masses derived from the 870 $\mu$m continuum flux. This is a factor of a few higher than the typical ISM value of $\sim100$, which could indicate significant grain growth and inward radial drift. However, given the continuum brightness of our sources it is also very likely that their continuum emission is optically thick, in which case the gas-to-dust mass ratio here represents an upper limit.

Also shown in Figure \ref{fig: mgas versus mdust} are the twenty class II disks from the AGE-PRO ALMA large program \citep{AGEPRO_I_overview}. Their gas masses and gas-to-dust ratios are discussed extensively in \cite{AGEPRO_V_gas_masses}. We include them here mostly to highlight that in terms of both dust and gas mass the MAPS, AGE-PRO, and exoALMA large programs sample different groups of objects (see \citealt{Oberg2021MAPS,AGEPRO_I_overview,Teague2025exoALMA}).

\subsection{Comparing with gas disk masses from kinematics}
\label{sec: comparing with kinematics}

\begin{figure}
    \centering
    \includegraphics[width=\columnwidth]{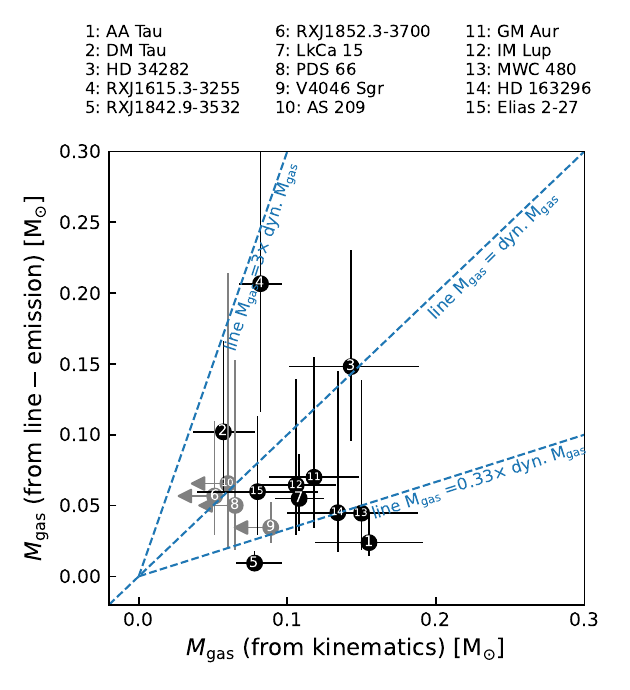}
    \caption{\label{fig: diff mgas vs mgas} A comparison of the emission line-based and kinematically derived gas disk masses. Kinematic gas mass upper limits are shown in gray. Blue lines show where line-based gas mass is 0.33$\times$, 1$\times$, and 3$\times$ the kinematically derived gas mass. }
\end{figure}

Figure \ref{fig: diff mgas vs mgas} shows a comparison of the line emission based gas masses from this work and kinematically measured gas masses from \cite{Veronesi2021, Martire2024, Longarini2025exoALMA} for 15 disks. Note that for five disks, AA Tau, RXJ1615.3-3255, RXJ1852.3-3700, PDS 66, and AS 209, the kinematically measured gas mass is an upper limit, set to $\mdisk/\mstar = 5\%$ (see \citealt{Martire2024,Longarini2025exoALMA} for details on these sources). For the five MAPS disks covered in \cite{Martire2024} we have adopted minimum gas mass uncertainties of 25\% of the measured gas disk mass as recommended by \cite{Andrews2024,Veronesi2024}. 

Overall the two methods agree with each other within their respective errors. Quantitatively, we find that both gas mass estimates agree within $1-2\sigma$ and lie within a factor 2-3 of each other. 
Interestingly we also see that the line based gas disk masses are consistently lower than the kinematically measured gas masses. Assuming that the kinematically measured gas masses represent the true disk mass, this suggests that the line based gas mass, or at least our method of deriving it, underestimates the true disk mass by a small factor. To quantify this factor we compute the median ratio using a Kaplan-Meier estimator. To incorporate the uncertainties on each gas mass we repeat this process 10000 times, each time drawing random line-emission and kinematically measured gas masses from their respective posterior distributions. We find that the dynamical masses in our sample are on average $2.3^{+0.7}_{-1.0}\times$ higher than the CO + \nhp\ gas masses.
We will discuss possible explanations for this in more detail in Section \ref{sec: interpreting the (dis)agreement between the mass measurement methods}. 

Three sources show large differences between their two gas disk mass estimates. RXJ1842.9-3532 and AA Tau have CO + \nhp\ gas masses that are $\sim5\times$ lower than their kinematically measured gas masses, while the CO + \nhp\ gas mass of RXJ1615.3-3255 is $\gtrsim 4\times$ larger than its kinematically measured one. For AA Tau the uncertainties on the gas mass from kinematics are large due to its high inclination and difficulties in extracting its thermal structure (see \citealt{Longarini2025exoALMA,GallowaySprietsma2025exoALMA}). As a result of this its two gas estimates are within $2\sigma$ of each other, similar to many of the other sources. 
For RXJ1842.9-3532 the uncertainties on both gas measurements are too small to explain their difference. From the kinematics side, a possible explanation could be that in the inner disk the $^{12}$CO and \xco\ emitting heights overlap, thereby complicating the extraction of the temperature structure. From the chemistry side, a low midplane ionization rate $(\zet \lesssim 10^{-19}\ \mathrm{s}^{-1})$ and/or an underabundance in N$_2$ could cause us to underestimate the CO + \nhp\ gas mass of RXJ1842.9-3532.
The difference for RXJ1615.3-3255 is harder to explain. As we will discuss in more detail in Section \ref{sec: interpreting the (dis)agreement between the mass measurement methods}, possible biases in the CO + \nhp\ gas mass measurement are more likely to underestimate the true disk mass. And kinematically this source does not show any obvious signs that could indicate that its kinematically measured mass is underestimated. It is worth mentioning that using CO + \nhp\ RXJ1615.3-3255 has a star-to-disk mass ratio of $\sim0.2$, which would have left a detectable kinematic signature (see \citealt{Andrews2024,Veronesi2024}).

\section{Discussion}
\label{sec: discussion}

\subsection{Interpreting the (dis)agreement between the two gas mass measurement methods}
\label{sec: interpreting the (dis)agreement between the mass measurement methods}

The previous section showed that emission line based and kinematically measured gas masses agree with each other, with the former typically being a factor $\sim2\times$ smaller than the latter. This is a promising result given that historically disk mass estimates could differ by up to two orders of magnitude (e.g. \citealt{Miotello2023}. However, we should keep in mind that both methods have their own set of systematic uncertainties. Here we highlight the most important ones for both methods and discuss how this affects the agreement in derived gas masses.

Starting with the kinematically measured gas masses, excellent summaries of the systematics were published by \cite{Andrews2024} and \cite{Veronesi2024}. In addition to observational effects such as spatial resolution bias, the main uncertainties for a kinematically measured gas mass are the temperature structure of the disk and the measurement of the emitting surface. The temperature is commonly measured from optically thick CO emission at one or more heights in the disk, depending on the number of available optically thick CO lines (e.g. \citealt{Law2021bMAPS,GallowaySprietsma2025exoALMA}), and must then be extrapolated to the midplane. 
Tests using both toy models and hydrodynamical simulations suggest that for massive disks $(\mdisk/\mstar \geq 0.1)$ the true disk mass can be recovered with little bias $(\lesssim20\%)$ and relatively low uncertainties $(\lesssim25\%)$ (see \citealt{Andrews2024,Veronesi2024} for details.)

Measuring the disk mass from line emission is more indirect than using kinematics because it, by necessity, requires the use of indirect tracers like CO as proxies for the bulk H$_2$. Most of the systematic uncertainties therefore arise from the assumptions made in the thermochemical models used to link the emission to the underlying gas mass. \cite{AGEPRO_V_gas_masses} provides a summary the main uncertainties for the method used in this work to measure line based gas masses, including the N$_2$ abundances, midplane ionization rate, and model assumptions. The first two most likely lead to an underestimation of the gas mass, which could explain why the line-based gas masses are lower than the kinematically measured ones. We will discuss the role of the cosmic ray ionization rate in more detail in Section \ref{sec: constraining midplane ionization rates}. As for the model assumption, \cite{AGEPRO_III_Lupus} showed that different thermochemical models seem to converge when it comes to the derived gas mass, particularly for the larger, more massive disks.

The vertical temperature structure of the disk can also have an important effect on line fluxes. \cite{QiWilner2024} showed that the presence of a thick vertical isothermal layer causes the CO column density to drop substantially beyond the CO iceline, resulting in a narrow ring of \nhp\ emission and lower integrated CO isotopologue fluxes (see also \citealt{Qi2019}). The presence of such a layer could therefore explain the difference between the line based and kinematically measured gas masses. However, \citealt{Qi2019} also showed that thick vertically isothermal layers are not present in three of their six disk sample, suggesting that it might not be a universal feature of protoplanetary disks.

\begin{figure}[!htb]
    \centering
    \includegraphics[width=\columnwidth]{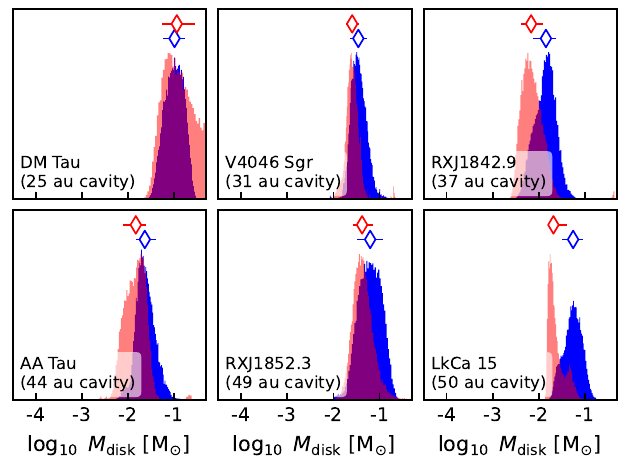}
    \caption{\label{fig: cavity gas masses} Comparison of the gas mass posteriors obtained using models with (in red) and without (in blue) inner dust cavities. Diamonds and lines above the distribution shows its median value and 16$^{\rm th}$,84$^{\rm th}$ quantiles. }
\end{figure}

To summarize, the kinematically measured gas mass provides a more direct measure of the true disk mass. However, it requires the disk to be massive relative to its star ($\mgas/\mstar \geq 0.05$), at intermediate inclination and overall axi-symmetric. Furthermore, measuring the disk mass requires a good understanding of the disk temperature structure and the line emitting height. Conversely, the CO + \nhp\ based gas mass measurement technique is more broadly applicable as it only requires integrated fluxes and can therefore be applied to a large sample, but it is more indirect and relies on our understanding of CO and \nhp\ disk chemistry.
It is therefore encouraging to see that both gas mass measurement techniques lie within a small factor ($\sim 2-3\times$) of each other.

\subsection{The effect of an inner dust cavity on measuring gas masses}
\label{sec: effect inner cavity}

As can be seen in Figure \ref{fig: B7 obs gallery}, a large number of the disks in this sample have large inner dust cavities (see also \citealt{Curone2025exoALMA}). These cavities are not included in the full disk thermochemical models of \citealt{AGEPRO_V_gas_masses}, but their presence will have an effect on the temperature structure of the disk (e.g. \citealt{Bruderer2013,Leemker2022}) which could affect the gas mass measurement. 
To quantify the effect of a dust cavity on our gas masses  we ran an additional set of models following the same setup as in \cite{AGEPRO_V_gas_masses}, but with inner dust cavities of 10, 30, and 60 au. These cavities are included as in e.g. \cite{Bruderer2012,vdMarel2015}, with the dust density inside the cavity set to zero. We do also include a small inner dust disk with a radius of 3 au and a dust depletion factor of $10^{-2}$, in line with the typical inner dust disks found in the survey of \citealt{FrancisvdMarel2020}. As it is computationally impracticable to rerun the full model grid we only run the dust cavity models for $\mdisk \in \left[10^{-4}, 10^{-3}, 10^{-2}, 10^{-1}\right]\ \msun$, $L_* = \left[0.25, 0.5, 1.0\right] \Lsun$, $\rc \in \left[30, 60, 120, 180 \right] \mathrm{au}$.
As the maximum stellar luminosity of these models is not representative of the Herbigs in our sample we will limit our analysis to the T-Tauri disks with dust cavities. 

Using these dust cavity models we re-derive \mgas\ for six T-Tauri disks with observed dust cavities: AA Tau, DM Tau, LkCa 15, RXJ1842.9-3532, RXJ1852.3-3700, and V4046 Sgr. These \mgas\ are measured in the same fashion as is outlined in Section \ref{sec: methods} except that the dust cavity size is now included as a free parameter, for which we assume a Gaussian prior with a mean equal to dust cavity size from \cite{FrancisvdMarel2020} and a standard deviation of 10 au ($\sim0.5\times$ the typical beam of the observations used to measure the dust cavity size). Figure \ref{fig: cavity gas masses} shows the resulting gas mass posterior distributions and compares them to the ones obtained using the full disk models. In all cases the gas mass decreases when a dust cavity is included. This is unsurprising as the presence of the cavity increases the overall disk temperature and therefore the line brightness of the CO lines, meaning that the observations can be reproduced with a lower gas mass. However, Figure \ref{fig: cavity gas masses} also shows that this effect is typically small: For the six disks examined there the reduction in gas mass is within a factor $\sim2.5$. 

\subsection{Constraining midplane ionization rates}
\label{sec: constraining midplane ionization rates}

\begin{figure}[!htb]
    \centering
    \includegraphics[width=\columnwidth]{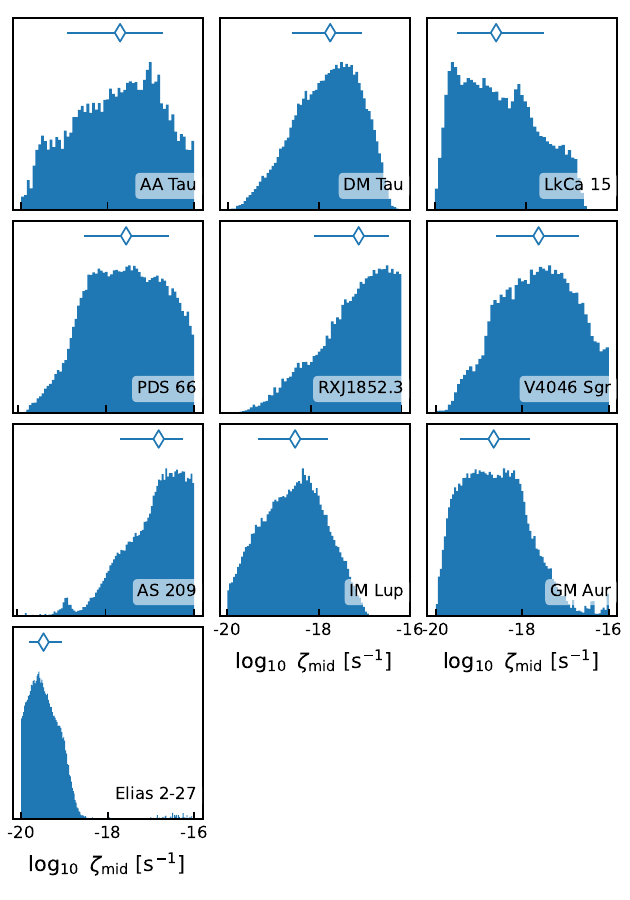}
    \caption{\label{fig: derived CR rates} Posterior distributions of the cosmic ray ionization rate (\zet), obtained by refitting the observation using the kinematically measured gas mass and its uncertainties as a prior. Diamonds and lines above the distribution shows it median value and 16$^{\rm th}$,84$^{\rm th}$ quantiles. 
    }
\end{figure}

As an ion, the abundance of \nhp\ depends on the midplane ionization rate, which is relatively poorly understood in protoplanetary disks (e.g. \citealt{Cleeves2015,Seifert2021,Aikawa2022MAPS,Long2024}). It is therefore one of the main sources of uncertainty for a disk mass obtained from CO isotopologues and \nhp. However, given our additional constraints on the gas mass from kinematics, we can also exploit this dependence and use \nhp\ to constrain the midplane ionization rate. We note here that ionization in the midplane of our models is set with a single ionization rate. In the disk midplane cosmic rays are thought to be the dominant source of ionization, but other sources, e.g. decay of radionuclides, scattered X-rays, can also contribute or even dominate if e.g. the midplane is shielded (e.g. \citealt{Cleeves2015}). The midplane ionization rate discussed here should therefore be interpreted as the sum of these processes.
It should also be noted that we are keeping the N$_2$ abundance fixed, effectively assuming that it is the main Nitrogen carrier in the disk. If this is not the case and the N$_2$ abundance is lower then the midplane ionization rates derived here would have to be higher to reproduce the observed amount of \nhp.  

Figure \ref{fig: derived CR rates} shows the posterior distributions of \zet\ where we have refit the observations but now use the kinematically derived gas masses and their uncertainties as a prior during the fit. We exclude RXJ1842.9-3532 and RXJ1615.3-3255 as their line-based and kinematic gas mass proved too different from each other to use the kinematic gas mass as a prior while fitting the line emission.
Overall we find a wide range of midplane ionization rates, from $\zet\approx10^{-17}\ \mathrm{s}^{-1}$ for AS 209 and RXJ1852.3-3700 to $\zet\approx 3\times10^{-20}\ \mathrm{s}^{-1}$ for Elias 2-27. We find a median ionization rate of $\zet = 1.5\times10^{-18}\ \mathrm{s}^{-1}$. However, the uncertainty of \zet\ for individual sources is large, typically $\sim1.5$ dex, suggesting that \nhp\ alone only provides a moderate constrain on the ionization rate. 

Comparing our rough, disk-averaged rates to previous studies that used more complex models and observations of multiple ions (e.g. \citealt{Seifert2021,Aikawa2022MAPS,Long2024}) we find overall agreement. For example, \cite{Long2024} find a moderate ionization rate of $10^{-18}\ \mathrm{s}^{-1}$ for DM Tau which agrees very well with our median value for the source $(\zet=1.5\times10^{-18}\ \mathrm{s}^{-1})$. Similarly, \cite{Aikawa2022MAPS} report midplane ionization rates $(\zeta_{\rm mid})$ of $\zeta_{\rm mid} \gtrsim 10^{-18}\ \mathrm{s}^{-1}$ for AS 209 and IM Lup, and $\zeta_{\rm mid} \lesssim 10^{-18}\ \mathrm{s}^{-1}$ for GM Aur. For AS 209 and GM Aur our results are in agreement with \citet{Aikawa2022MAPS}, but for IM Lup we find a somewhat lower value of $\sim3\times10^{-19}\ \mathrm{s}^{-1}$. However, \cite{Seifert2021} showed that IM Lup has radially varying midplane ionization rate, with $\zeta_{\rm mid} \lesssim 10^{-20}\ \mathrm{s}^{-1}$ inside 100 au and $\zeta_{\rm mid} \gtrsim 10^{-17}\ \mathrm{s}^{-1}$ between 100 and 300 au. As our sole constraint comes from the integrated \nhp\ flux we are probing a mix of these two regions, which could explain our overall lower midplane ionization rate.

\subsection{The role of \nhp\ in constraining Herbig gas disk masses}
\label{sec: n2hp for herbigs}

\begin{figure}[!htb]
    \centering
    \includegraphics[width=\columnwidth]{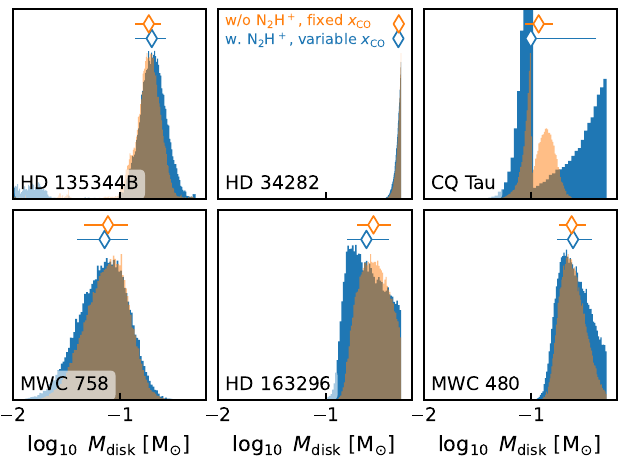}
    \caption{\label{fig: Herbig with N2Hp} Comparison of the gas mass posteriors for the six Herbig disks for cases where \nhp\ is excluded from the fit and \abu\ kept fixed (in orange) and where \nhp\ is included in the fit and \abu\ is a free parameter (in blue). Diamonds and lines above the distribution shows it median value and 16$^{\rm th}$,84$^{\rm th}$ quantiles. }
\end{figure}

In this work we have measured gas masses for T-Tauri and Herbig disks in slightly different ways, the largest difference being that for the Herbig disks we do not fit the \nhp\ observations. Initial gas measurements using CO found that the gas-to-dust mass ratios of T-Tauri disks were much lower than the interstellar medium value of  $\sim100$ (e.g. \citealt{WilliamsBest2014,miotello2017,Long2017}). Comparisons with other gas mass traces such as HD \citep{Bergin2013,McClure2016} suggested that the CO is underabundant in the warm molecular layer of protoplanetary disks, leading to the inclusion of \nhp\ as an additional constraint on the gas-phase CO abundance \citep{Anderson2019,Anderson2022,Trapman2022b,AGEPRO_V_gas_masses}. For Herbig disks CO-based gas-to-dust mass ratios are routinely found to be at or above the ISM value of $\sim100$ (e.g. \citealt{Stapper2024}). This suggests that the processes that reduce the gas-phase CO abundance in T-Tauri disks do not significantly affect disks around Herbig stars.

However, this does not exclude the possibility that the gas-phase CO in Herbig disks could be at least mildly underabundant and that their CO-based gas masses could therefore underestimate their true gas masses. Indeed, several studies have found that at least some Herbig disks seem underabundant in CO to some degree (e.g. \citealt{Zhang2021MAPS, PanequeCarreno2024,Rosotti_exoALMA}). We can test this hypothesis by fitting the \nhp\ flux in addition to the CO fluxes, identical to how we fitted the T-Tauri disks (see Section \ref{sec: methods}). Unfortunately the models in \citet{Stapper2024} do not include synthetic observations of \nhp. We therefore ran another set of models analogous to the new $L_* = 3\ \Lsun$ models described in Section \ref{sec: methods} but using the observed stellar spectrum of HD 163296 ($L_* = 17\ \Lsun$; \citealt{Dionatos2019}). 

Using these models we fit the observations of the six Herbig disks in our sample twice, once where we fit the 1.3 mm continuum and \xco,  \cyo, and \nhp\ line fluxes while having the peak CO abundance be a free parameter and once where we fit the 1.3 mm continuum and \xco,  \cyo\ line fluxes and keep the peak CO abundance fixed (analogous to fitting the Herbig disks using the \citet{Stapper2024} models). We note the \citet{Stapper2024} models cover a wider range of parameters relevant for Herbig disks (e.g, $L_*$, scale height) and have a higher mass resolution that our models. The disk masses derived here do therefore not supersede the ones presented in Section \ref{sec: line based gas disk masses}.   

Figure \ref{fig: Herbig with N2Hp} shows the resulting gas mass posteriors for the two fits. For all six source we find that the gas masses are near identical, with the mean estimates lying within $\sim15\%$ of each other.  The dual peak structure of the posterior of CQ Tau is due to a lack of models. Its best fit characteristic radius using the \citet{Stapper2024} models is $\rc \approx 12$ au, while the smallest model in our grid with $\mdisk = 0.5~\msun$ has $\rc=15$ au. This causes the MCMC to oscillate between these models and models with $\mdisk = 0.1~\msun, \rc=5$ au, which are included in the grid. Due to this incomplete fit we caution against using the $\abu$ for CQ Tau and we have excluded it from our analysis here.

\begin{figure}[!htb]
    \centering
    \includegraphics[width=\columnwidth]{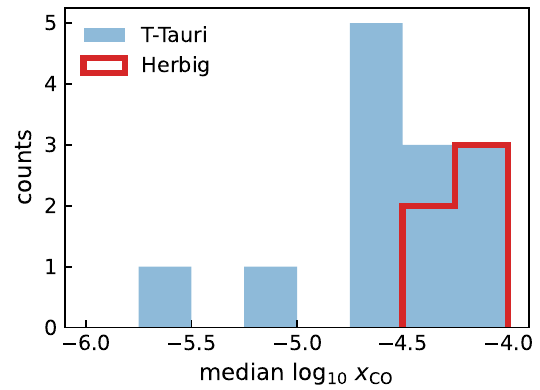}
    \caption{\label{fig: median CO abundances} 
    Histogram of the median peak CO abundances, split between the Herbig disks (in red) and the T-Tauri disks (in blue).}
\end{figure}

Figure \ref{fig: median CO abundances} shows a histogram of the median peak CO abundances of the Herbig disks discussed here and the T-Tauri disks examined previously. For the Herbig disks we find median values of $\log_{10} \abu \approx -4.35 - 4.1$, which is very close to the ISM value of $\log_{10} \abu \approx -4$. Peak CO abundances of the T-Tauri disks are typically lower, peaking at $\log_{10} \abu \approx -4.6$ and they can be as low as $\log_{10} \abu = -5.6$ (Table \ref{tab: gas masses sample}).

This suggests different chemical paths for the T-Tauri and Herbig disks, where in the colder T-Tauri disks a significant fraction of the gaseous CO freezes out and is converted into less volatile species such as CO$_2$, whereas in the warmer Herbig disks most of the CO stays in the gas and is not chemically converted (see \citealt{Oberg2023} for a recent review).

The high peak CO abundances found here seem at odds with the lower values found in other works (e.g. \citealt{Zhang2021MAPS, PanequeCarreno2024,Rosotti_exoALMA}). \cite{PanequeCarreno2024} and \cite{Rosotti_exoALMA} independently showed that the emitting height of optically thick CO lines are strongly correlated to both the disk mass and the CO abundance. They found that in order to reproduce the CO emitting height of a number of Herbig disks using the gas mass from previous estimates they need reduce the CO abundance by a factor of up to ten. 

One possible explanation of the dichotomy between these works and the results here is that the two methods trace different parts of the disk. The \nhp~(3-2) emission predominantly traces the layer between the CO and N$_2$ icelines (e.g. \citealt{Qi2013,Qi2019,vantHoff2017}) and is therefore most sensitive to the CO abundance of the gas close to the CO iceline. Conversely, the CO emitting height is most sensitive to the CO abundance at the top of the CO emitting layer, close to where CO is photo-dissociated. 

We should note here the these results apply specifically for Herbig disks. Similar tests for T-Tauri disks show that their gas masses can be $\sim3-30\times$ higher and their CO peak gas abundances $\sim3-30\times$ lower if \nhp\ is included with CO when measuring gas masses (e.g. Table \ref{tab: gas masses sample}; see also \citealt{Anderson2019,Anderson2022,Trapman2022,Sturm2023}).

\section{Conclusions}
\label{sec: conclusions}

In this work we have presented new ALMA Band 7 observations of \nhp~(3-2) for 11 protoplanetary disks and Band 6 \cyo~(2-1) observations for one disk. These we combine with archival \nhp, \xco, and \cyo\ observations to measure gas disk masses for 19 disks, primarily from the exoALMA and MAPS ALMA Large programs. For the 15 disks where kinematically measured gas masses are available we compare the two gas mass measuring techniques. Our main conclusions are listed below

\begin{itemize}
    \item For T-Tauri disks we find clear correlations between the continuum flux and the \cyo\ and \nhp\ line fluxes similar to what previously found in the literature. Disks around Herbig stars lie on the same \cyo-continuum flux correlation, but have 1-2 orders of magnitude lower \nhp\ fluxes compared to T-Tauri disks with similar continuum fluxes. This is likely due to the higher overall temperature and lower CO freeze-out in these disks, resulting in less CO-poor gas and therefore lower overall \nhp\ abundances.
    \item We find overall high gas masses between $\sim0.01-0.1\ \msun$ and gas-to-dust mass ratios in the range $\sim80-530$ for our sources. However, these high gas-to-dust ratios are likely lower upper limits due the continuum emission used to derive dust masses being optically thick. 
    \item There is good agreement between the line emission based gas masses from this work and the ones measured from gas kinematics in the literature. With two exceptions, RXJ1842.9-3532 and RXJ1615.3-3255, the gas estimates from the two methods lie within a factor $\sim3$ ($1-2\sigma$) from each other.
    \item Gas disk masses from CO + \nhp\ are on average a factor $2.3^{+0.7}_{-1.0}\times$ lower than the kinematically measured gas masses. Assuming the kinematics trace the true disk mass, this could point towards slightly lower N$_2$ abundances and/or lower midplane ionization rates than typically assumed.
    \item Herbig disks are found to have ISM level CO gas abundances based on their CO and \nhp\ fluxes, which sets them apart from T-Tauri disks where the CO abundance in the CO emitting layer is typically $\sim3-30\times$ lower. 
\end{itemize}

The agreement between the gas masses measured from gas kinematics and CO + \nhp\ is promising as the former provides direct estimate of the mass that does not depend on disk chemistry, whereas the latter is much broadly applicable. It shows that multi-molecule line fluxes are a robust tool to accurately measure disk masses at least for extended disks.

\section*{acknowledgements}
We would like to thank the anonymous referee for their insightful feedback.
LT and KZ acknowledge the support of the NSF AAG grant \#2205617. 
CL has received funding from the European Union's Horizon 2020 research and innovation program under the Marie Sklodowska-Curie grant agreement No. 823823 (DUSTBUSTERS) and by the UK Science and Technology research Council (STFC) via the consolidated grant ST/W000997/1.
GR acknowledges funding from the Fondazione Cariplo, grant no. 2022-1217, and the European Research Council (ERC) under the European Union's Horizon Europe Research \& Innovation Programme under grant agreement no. 101039651 (DiscEvol). Views and opinions expressed are however those of the author(s) only, and do not necessarily reflect those of the European Union or the European Research Council Executive Agency. Neither the European Union nor the granting authority can be held responsible for them.
JB acknowledges support from NASA XRP grant No. 80NSSC23K1312.
MB, DF, and JS have received funding from the European Research Council (ERC) under the European Union’s Horizon 2020 research and innovation programme (PROTOPLANETS, grant agreement No. 101002188). Computations have been done on the ’Mesocentre SIGAMM’ machine,hosted by Observatoire de la Cote d’Azur.
PC acknowledges support by the Italian Ministero dell'Istruzione, Universit\`a e Ricerca through the grant Progetti Premiali 2012 – iALMA (CUP C52I13000140001) and by the ANID BASAL project FB210003.
S.F. is funded by the European Union (ERC, UNVEIL, 101076613). Views and opinions expressed are however those of the author(s) only and do not necessarily reflect those of the European Union or the European Research Council. Neither the European Union nor the granting authority can be held responsible for them. S.F. acknowledges financial contribution from PRIN-MUR 2022YP5ACE.
MF is supported by a Grant-in-Aid from the Japan Society for the Promotion of Science (KAKENHI: No. JP22H01274).
JDI acknowledges support from an STFC Ernest Rutherford Fellowship (ST/W004119/1) and a University Academic Fellowship from the University of Leeds.
Support for AFI was provided by NASA through the NASA Hubble Fellowship grant No. HST-HF2-51532.001-A awarded by the Space Telescope Science Institute, which is operated by the Association of Universities for Research in Astronomy, Inc., for NASA, under contract NAS5-26555.
GL acknowledges support from the European Union’s Horizon 2020 research and innovation programme under the Marie Sklodowska-Curie grant agreement \#823823 (RISE DUSTBUSTERS project) and from the Italian MUR through PRIN 20228JPA3A. 
C.P. acknowledges Australian Research Council funding  via FT170100040, DP18010423, DP220103767, and DP240103290.
H.-W.Y.\ acknowledges support from National Science and Technology Council (NSTC) in Taiwan through grant NSTC 110-2628-M-001-003-MY3 and from the Academia Sinica Career Development Award (AS-CDA-111-M03).
Support for BZ was provided by The Brinson Foundation.
CH acknowledges support from the National Science Foundation Astronomy and Astrophysics Research Grants program No. 2407679.
AJW has received funding from the European Union’s Horizon 2020
research and innovation programme under the Marie Skłodowska-Curie grant
agreement No 101104656.

This paper makes use of the following ALMA data: ADS/JAO.ALMA\#2023.1.00334.S, \#2022.1.00485.S, 
\#2021.1.01123.L,
\#2012.1.00158.S,
\#2012.1.00870.S,
\#2015.1.00192.S,
\#2015.1.00678.S, 
\#2015.1.01017.S,
\#2016.1.00484.L,
\#2016.1.00724.S,
\#2017.1.00069.S,
\#2017.1.00940.S,
\#2017.1.01404.S,
\#2017.1.01419.S,
\#2018.1.00689.S,
\#2018.1.00945.S,
\#2018.1.01055.L,
\#2019.1.01683.S
ALMA is a partnership of ESO (representing its member states), NSF (USA) and NINS (Japan),
together with NRC (Canada), MOST and ASIAA (Taiwan), and KASI (Republic of Korea), in
cooperation with the Republic of Chile. The Joint ALMA Observatory is operated by
ESO, AUI/NRAO and NAOJ. The National Radio Astronomy Observatory is a facility of the National
Science Foundation operated under cooperative agreement by Associated
Universities, Inc.

All figures were generated with the \texttt{PYTHON}-based package \texttt{MATPLOTLIB} \citep{Hunter2007}. This research made use of Astropy,\footnote{http://www.astropy.org} a community-developed core Python package for Astronomy \citep{astropy:2013, astropy:2018}.

\facility{ALMA}

\software{CASA \citep{McMullin2007,CASAteam2022}, Astropy \citep{astropy:2013,astropy:2018}, MATPLOTLIB \citep{Hunter2007}}

\bibliographystyle{aasjournal}
\bibliography{references}

\begin{appendix}
\vspace{-0.5cm}

\section{Observational details}
\label{sec: obs details}
\vspace{-0.5cm}

\begin{table*}[htb]
\centering
\caption{\label{tab: observational info} ALMA observation details }
\def\arraystretch{1.0}
\begin{tabular*}{0.96\textwidth}{lc|cccccc}
\hline\hline
& \multicolumn{7}{c}{ALMA Band 7 - N$_2$H$^+$}\\
Name & Date & Antennas & Baselines & Time on source & Bandpass Calibrator & Phase Calibrator & Flux Calibrator\\
     &      &          & [m]       & [min]          &               &            &          \\
\hline
 RXJ1615.3-3255 & 2 April 2024 & 42 & 15-251 & 8 & J1924-2914 & J1626-2951 & J1924-2914 \\
  & 1 July 2024 & 41 & 15-1261 & 22 & J1427-4206 & J1427-4206 & J1626-2951 \\
  & 1 July 2024 & 36 & 15-649 & 23 & J1427-4206 & J1610-3958 & J1427-4206 \\
 RXJ1852.3-3700 & 1 April 2024 & 41 & 15-280 & 8 & J1924-2914 & J1839-3453 & J1924-2914 \\
  & 28 May 2024 & 42 & 15-984 & 23 & J1924-2914 & J1802-3940 & J1924-2914 \\
 RXJ1842.9-3532 & 1 April 2024 & 41 & 15-280 & 8 & J1924-2914 & J1839-3453 & J1924-2914 \\
  & 28 May 2024 & 42 & 15-984 & 23 & J1924-2914 & J1802-3940 & J1924-2914 \\
 AA Tau & 27 July 2024 & 46 & 14-499 & 9 & J0423-0120 & J0438+3004 & J0423-0120 \\
  & 8 Sep 2024 & 43 & 15-1604 & 25 & J0423-0120 & J0438+3004 & J0423-0120 \\
 PDS 66 & 21 March 2024 & 45 & 15-313 & 9 & J1427-4206 & J1147-6753 & J1427-4206 \\
  & 29 May 2024 & 41 & 15-984 & 25 & J1427-4206 & J1147-6753 & J1427-4206 \\
 HD 143006 & 12 April 2024 & 44 & 15-314 & 8 & J1924-2914 & J1625-2527 & J1924-2914 \\
  & 4 Jun 2024 & 40 & 15-1397 & 22 & J1924-2914 & J1551-1755 & J1924-2914 \\
 HD 135344B & 12 April 2024 & 44 & 15-313 & 8 & J1924-2914 & J1457-3539 & J1924-2914 \\
  & 28 May 2024 & 41 & 15-984 & 23 & J1256-0547 & J1457-3539 & J1256-0547 \\
 CQ Tau & 31 July 2024 & 45 & 14-483 & 9 & J0423-0120 & J0521+1638 & J0423-0120 \\
  & 22 Sep 2024 & 43 & 15-1397 & 25 & J0423-0120 & J0521+1638 & J0423-0120 \\
 HD 34282 & 1 Aug 2024 & 46 & 14-313 & 23 & J0423-0120 & J0501-0159 & J0423-0120 \\
  & 1 Nov 2024 & 46 & 14-313 & 8 & J0423-0120 & J0501-0159 & J0423-0120 \\
 MWC 758 & 31 July 2024 & 45 & 14-483 & 9 & J0423-0120 & J0521+1638 & J0423-0120 \\
  & 22 Sep 2024 & 43 & 15-1397 & 25 & J0423-0120 & J0521+1638 & J0423-0120 \\
 Elias 2-27 & 6 Oct 2022 & 10 & 8-48 & 19 & J1427-4206 & J1625-2527 & J1427-4206 \\
  & 7 Oct 2022 & 10 & 8-48 & 34 & J1427-4206 & J1625-2527 & J1427-4206 \\
  & 9 Oct 2022 & 10 & 8-48 & 34 & J1427-4206 & J1625-2527 & J1427-4206 \\
  & 9 Jan 2023 & 40 & 15-740 & 28 & J1427-4206 & J1700-2610 & J1427-4206 \\
\hline\hline
\multicolumn{8}{c}{}\\
& \multicolumn{7}{c}{ALMA Band 6 - C$^{18}$O}\\
Name & Date & Antennas & Baselines & Time on source & Bandpass Cal. & Phase Cal. & Flux Cal.\\
     &      &          & [m]       & [min]          &               &            &          \\
\hline
 RXJ1842.9-3532 & 21 Dec 2023 & 46 & 15-1397 & 32 & J1924-2914 & J1826-3650 & J1924-2914 \\
  & 25 March 2024 & 48 & 15-314 & 9 & J1924-2914 & J1826-3650 & J1924-291 \\
\hline\hline
\end{tabular*}
\begin{minipage}{0.93\textwidth}
\vspace{0.1cm}
{\footnotesize{\textbf{Notes:}
}}
\end{minipage}
\end{table*}

\begin{table*}[htb]
\centering
\caption{\label{tab: fluxes and programs} ALMA observations \& integrated fluxes}
\def\arraystretch{1.0}
\begin{tabular*}{0.95\textwidth}{l|cccccc}
\hline\hline
source & \multicolumn{4}{c}{Disk-integrated fluxes (mJy [km s$^{-1}$])} & Program IDs & refs\\
       & 285 GHz cont. & $^{13}$CO~(2-1) & C$^{18}$O~(2-1) & N$_2$H$^+$~(3-2) & & \\
\hline 
DM Tau     & $99.1 \pm 1.4$   & $5341.7 \pm 34.3$ & $1275.6 \pm 21.8$ & $1437.6 \pm 39.2$ & 2016.1.00724.S, 2015.1.00678.S & [3,2,2,3] \\
V4046 Sgr  & $576.0 \pm 12.0$ & $8275.7 \pm 44.3$ & $1936.0 \pm 38.5$ & $3736.1 \pm 31.6$ & 2016.1.00724.S, 2015.1.00678.S & [3,2,2,3] \\
RXJ1852.3-3700  & $93.0 \pm 0.6$   & $1587.5 \pm 37.7$ & $504.4 \pm 22.1$ & $541.0 \pm 17.1$ & 2018.1.00689.S, 2023.1.00334.S & [1,1,1,1]\\
RXJ1842.9$\dagger$ &  $87.6 \pm 0.5$ & - & $238 \pm 4$ & $359.4 \pm 7.9$ & 2023.1.00334.S, 2021.1.01123.L & [1,4,1,1] \\
LkCa 15    & $281.0 \pm 6.9$  & $6074.2 \pm 16.1$ & $1198.1 \pm 15.6$ & $1869.0 \pm 33.5$ & 2018.1.00945.S, 2015.1.00678.S & [3,5,5,3] \\
PDS 66     & $162.2 \pm 8.0$  & $790.0 \pm 50.0$  & $210.0 \pm 20.0$ & $1000.7 \pm 14.5
$ & 2017.1.01419.S, 2023.1.00334.S & [1,7,7,1] \\
CQ Tau     & $207.5 \pm 1.9$  & $1120.0 \pm 40.0$ & $600.0 \pm 20.0$ & $19.9 \pm 5.88$ & 2017.1.01404.S, 2023.1.00334.S& [1,8,8,1]\\
HD 34282   & $205.2 \pm 6.9$ & $3460.0 \pm 50.0$ & $1890.0 \pm 40.0$ & $288.2 \pm 9.2$ & 2017.1.01404.S, 2023.1.00334.S& [1,8,8,1]\\
MWC 758    & $101.3 \pm 0.4$  & $1870.0 \pm 70.0$ & $780.0 \pm 40.0$ & $71.8 \pm 6.3$ & 2017.1.00940.S, 2023.1.00334.S& [1,8,8,1]\\
GM Aur     & $286.4 \pm 2.6$  & $5028.0 \pm 48.0$ & $1092.0 \pm 39.0$ & $1484.0 \pm 14.6$ & 2018.1.01055.L, 2015.1.00678.S & [3,9,9,3]\\
AS 209     & $289.0 \pm 4.5$  & $2269.0 \pm 38.0$ & $538.0 \pm 27.0$ & $884.5 \pm 14.3$ & 2018.1.01055.L, 2015.1.00678.S & [3,9,9,3]\\
IM Lup     & $243.7 \pm 6.2$  & $8370.0 \pm 83.0$ & $1592.0 \pm 60.0$ & $2040.0 \pm 49.0$ & 2018.1.01055.L, 2015.1.00678.S & [3,9,9,3]\\
MWC 480    & $490.0 \pm 49.0$ & $8361.0 \pm 57.0$ & $3017.0 \pm 41.0$ & $228.0 \pm 13.0$ & 2018.1.01055.L, 2015.1.00657.S & [3,9,9,10]\\
HD 163296  & $791.6 \pm 1.6$  & $15885.0 \pm 80.0$ & $5783.0 \pm 51.0$ & $556.1 \pm 49.1$ & 2018.1.01055.L, 2012.1.00681.S & [3,9,9,11]\\
\hline
       & 285 GHz cont. & $^{13}$CO~(3-2) & C$^{18}$O~(3-2) & N$_2$H$^+$~(3-2) & & \\
\hline
RXJ1615.3-3255  & $245.9 \pm 0.4$ & $11504.0 \pm 41.8$ & $3226.8 \pm 48.2$ & $563.6 \pm 6.2$ & 2012.1.00870.S, 2023.1.00334.S & [1,1,1,1] \\
RXJ1842.9-3532$\dagger$ &  -  & $2831.2 \pm 10.3$ & - &  & 2023.1.00334.S, 2021.1.01123.L & [1,4,1,1] \\
AA Tau     & $112.0 \pm 0.5$ & $5102.0 \pm 42.0$ & $1217.5 \pm 53.4$ & $686.6 \pm 10.8$ & 2015.1.01017.S, 2023.1.00334.S& [1,7,7,1]\\
HD 143006  & $102.0 \pm 0.4$ & $1668.0 \pm 7.6$ & $483.3\pm 10.1$ & $330.9 \pm 8.8$ & 2019.1.01683.S, 2023.1.00334.S& [1,1,1,1]\\
HD 135344B & $302.8 \pm 0.6$ & $7482.5 \pm 39.1$ & $3115.4 \pm 46.1$ & $40.3 \pm 5.8$ & 2012.1.00158.S, 2023.1.00334.S& [1,12,12,1]\\
Elias 2-27 & $467.6 \pm 0.5$ & $22260.0 \pm 90.0$ & $4450.0 \pm 60.0$ & $447 \pm 6.5$ & 2016.1.00724.S, 2022.1.00485.S& [1,13,13,1]\\
\hline\hline
\end{tabular*}
\begin{minipage}{0.95\textwidth}
\vspace{0.1cm}
{\footnotesize{\textbf{Notes:}  $^{\dagger}$: For RXJ1842.9 the two CO isotopologue lines are \xco~(3-2) from \cite{GallowaySprietsma2025exoALMA} and \cyo~(2-1) from this work. Flux uncertainties do no include absolute flux uncertainties. References are given for each individual line. 
[1]: \emph{This work}, 
[2]: \citealt{Flaherty2020}, 
[3]: \citealt{Qi2019},
[4]: \citealt{GallowaySprietsma2025exoALMA},
[5]: \citealt{Sturm2023},
[6]: \citealt{Loomis2017},
[7]: \citealt{Ribas2023},
[8]: \citealt{Stapper2024},
[9]: \citealt{Oberg2021MAPS},
[10]: \citealt{Loomis2020},
[11]: \citealt{Qi2015},
[12]: \citealt{vdMarel2015},
[13]: \citealt{Paneque-Carreno2023}}}
\end{minipage}
\end{table*}

\end{appendix}

\end{document}